\documentclass[11pt,english,twoside]{article}

\usepackage[T1]{fontenc}
\usepackage[latin1]{inputenc}
\usepackage[english]{babel}
\usepackage{lmodern}
\usepackage{amsmath,amssymb}
\usepackage{float}
\usepackage{graphicx}
\usepackage[dvips]{epsfig}
\usepackage{psfrag}
\usepackage{hyperref}

\usepackage{lscape}
\usepackage[all]{xy}

\newcommand{\beq}{\begin{equation}}
\newcommand{\eeq}{\end{equation}}
\newcommand{\bea}{\begin{eqnarray}}
\newcommand{\eea}{\end{eqnarray}}
\newcommand{\nn}{\nonumber}
\newcommand{\w}{\wedge}
\newcommand{\del}{\partial}
\newcommand{\aaa}{{\cal A}}
\newcommand{\bbb}{{\cal B}}

\newcommand{\eee}{{\cal E}}
\newcommand{\fff}{{\cal F}}
\newcommand{\mmm}{{\cal M}}
\newcommand{\nnn}{{\cal N}}

\newcommand{\ov}{\overline}

\newcommand{\Gg}{\mathfrak{g}}

\newcommand{\Ga}{\mathfrak{a}}
\newcommand{\Gb}{\mathfrak{b}}
\newcommand{\Gc}{\mathfrak{c}}
\newcommand{\Gd}{\mathfrak{d}}

\def\d {{\rm d}}

\def\di {d_i}
\def\dg {d_{{\rm g}}}
\def\ap {\alpha^{\prime}}
\def\apd {\alpha^{\prime \ \frac{1}{2}}}
\def\appd {\alpha^{\prime \ \frac{3}{2}}}
\def\a {\alpha^{\prime \ 0}}
\def\app {\alpha^{\prime \ 2}}

\def\tr {{\rm tr}}
\def\tD {\tilde{D}}
\def\tg {\tilde{g}}
\def\tB {\tilde{B}}
\def\tH {\tilde{H}}
\def\tp {\tilde{\phi}}
\def\tR {\tilde{R}}
\def\tM {\tilde{M}}
\def\tN {\tilde{N}}
\def\tP {\tilde{P}}
\def\tQ {\tilde{Q}}
\def\to {\tilde{\omega}_+}
\def\ta {\tilde{a}}
\def\tb {\tilde{b}}
\def\tc {\tilde{c}}
\def\te {\tilde{e}}

\def\tG {\tilde{\Gamma}}
\def\tna {\tilde{\nabla}}
\def\al {\alpha}
\def\be {\beta}

\def\JN {J_{{\cal N}}}
\def\ON {\Omega_{{\cal N}}}

\DeclareMathOperator{\re}{Re}
\DeclareMathOperator{\im}{Im}

\makeatletter
\@addtoreset{equation}{section}
\makeatother

\begin{document}

\begin{titlepage}

\begin{center}

\rightline{\small LMU-ASC 05/11}

\vskip 3.2cm

\begin{LARGE}
   \textbf{Heterotic string from a higher dimensional perspective}
   \vskip 0.2cm
   \textbf{   }
   \vskip 0.3cm
  \textbf{ }
\end{LARGE}

\vskip 1.6cm

\textbf{David Andriot}

\vskip 0.6cm
\textit{Arnold Sommerfeld Center, c/o LS L\"ust, Ludwig-Maximilians-Universit\"at, \\
Theresienstrasse 37, 80333 M\"unchen, Germany}\\
\vskip 0.2cm
David.Andriot@physik.uni-muenchen.de

\end{center}

\vskip 2.1cm

\begin{center}
{\bf Abstract}
\end{center}

The (abelian bosonic) heterotic string effective action, equations of motion and Bianchi identity at order $\ap$ in ten dimensions, are shown to be equivalent
 to a higher dimensional action, its derived equations of motion and Bianchi identity. The two actions are the same up to the gauge fields: the latter are absorbed in
the higher dimensional fields and geometry. This construction is inspired by heterotic T-duality, which becomes natural in this higher dimensional theory.

We also prove the equivalence of the heterotic string supersymmetry conditions with higher dimensional geometric conditions. Finally,
some known K\"ahler and non-K\"ahler heterotic solutions are shown to be trivially related from this higher dimensional perspective, via a simple
 exchange of directions. This exchange can be encoded in a heterotic T-duality, and it may also lead to new solutions.



\vfill

\end{titlepage}

\tableofcontents

\setcounter{page}{0}

\newpage

\section{Introduction}

String theories are known to be related by a web of dualities. It is a common belief that they could be different partial descriptions of a more
fundamental theory. Such a theory would then admit these dualities as symmetries. To understand better the structure behind these different descriptions,
or at least, to have a better control on these dualities, it is of interest to construct and work with theories which are manifestly duality-covariant.
 Some progress in this direction has been made, at least at the level of the effective descriptions of string theories, in
 particular supergravities (SUGRA). A common feature of theories in which a duality transformation appears explicitly, possibly as
 a symmetry, is to consider additional dimensions. The duality is then often promoted to a simple geometric transformation involving these additional
 dimensions. F-theory is an example of such a construction for S-duality \cite{V}. For T-duality, several constructions have been worked out. A theory in
 which the target space fields are independent of $\di$ dimensions (these are then $\di$ isometries) admits transformations under the T-duality
 group $O(\di,\di)$. We will always mean here the continuous real group for the effective supergravity description, but it
 turns out to be broken to its discrete subgroup for the full string theory (for a review see \cite{GPR, T}). A theory dimensionally reduced along the $\di$
 isometries will then admit T-duality as a symmetry, but the latter would appear more explicitly by going back to a set-up with the $\di$ additional
 dimensions. Generalized Complex Geometry (GCG) \cite{HG, GMPW, K} or Doubled Geometry \cite{HuD} (and Double Field Theory \cite{DFT}) even consider spaces
 of doubled dimension\footnote{Note that the doubling is not the same: in Doubled Geometry, the doubled dimensions are independent of the initial ones,
 which is not the case in GCG.} $2\di$. This allows to give a geometric meaning to
 the T-duality group, which makes these transformations even more natural. In this paper, we are interested in heterotic string with a gauge group having
 a $\dg$-dimensional Cartan subgroup. In that case, the T-duality group is enhanced to $O(\di,\di+\dg)$. This was first noticed via lattice and moduli
 space studies \cite{N, NSW}, then via world-sheet \cite{GRV, SW, GR} and effective actions \cite{HS, MS} analysis. We will study here a
 geometrization of the $\di+\dg$ dimensions, which, to the best of our knowledge, has not been proposed so far. It involves non-trivially the gauge fields,
and put them on equal footing with the other fields. We will consider an associated higher dimensional theory and show it is equivalent to the (abelian bosonic) heterotic string effective
 description at order $\ap$. The heterotic T-duality is then more natural in this higher dimensional construction.
 
 The prospect of getting a theory
 explicitly covariant under the heterotic T-duality transformation, is the first motivation for this study. In particular, we come back in section \ref{secCcl} to
 the T-duality covariant rewriting of the (abelian bosonic) heterotic string effective action using a generalized metric, similarly to \cite{MS, DFT} for the NSNS action. Such a rewriting follows automatically from the higher dimensional theory we consider here. The equivalence shown will turn out to have further interesting consequences. For instance, it allows to relate in a trivial way some heterotic solutions which look a priori very different (the K\"ahler/non-K\"ahler solutions). Finally, note that the equivalent higher dimensional theory is very close to the effective description of bosonic string; we then comment in section \ref{secCcl} on possible relations of our (target space) equivalence to the (world-sheet) embedding of heterotic string into bosonic string.\\

 In order to motivate our construction, let us give more details on the heterotic T-duality. The action of the $O(\di,\di)$ group on the metric $g$ and
 the $B$-field $B$ can usually be encoded via the fractional linear transformation (detailed in (\ref{fraclin})), which is acting on the
 $\di \times \di$ matrix\footnote{The $m,n$ coefficients of this matrix are given by $g_{mn}+B_{mn}$, where those coefficients are
 obtained in a local coordinate basis of forms $\d x^m$.} $g+B$. Out of world-sheet analysis \cite{GRV, SW, GR}, it was pointed out that in the
heterotic case, one could use a $(\di+\dg) \times (\di+\dg)$ matrix to play the role of $g+B$. This bigger matrix could be decomposed in its symmetric and
antisymmetric part, giving a pseudo metric $\tg$ and a pseudo $B$-field $\tB$ on a bigger $(\di+\dg)$-dimensional space. Their components were given by
\beq
\tg=\begin{pmatrix} g_{mn}+g_{\Ga\Gb}A_m^{\Ga} A_n^{\Gb} & g_{\Gb \Ga} A^{\Ga}_m \\ g_{\Ga \Gb} A^{\Gb}_n & g_{\Ga \Gb} \end{pmatrix} \ , \
\tB=\begin{pmatrix} B_{mn} & g_{\Gb \Ga} A^{\Ga}_m \\ -g_{\Ga \Gb} A^{\Gb}_n & B_{\Ga \Gb} \end{pmatrix} \ , \label{pseudomatrix}
\eeq
where the connection $A^{\Ga}_m$ was proposed to be related to the heterotic gauge potential. On the $\dg$ part of the space, the metric $g_{\Ga \Gb}$,
 respectively the $B$-field $B_{\Ga\Gb}$, were given by a symmetrized, respectively antisymmetrized, version of the Cartan matrix of the gauge algebra,
 as already proposed in \cite{EGRS}. We can rewrite these pseudo fields in the following way
\bea
\d \tilde{s}^2&=&g_{mn} \d x^m \d x^n + g_{\Ga \Gb} (\d x^{\Ga}+ A^{\Ga}_m \d x^m)(\d x^{\Gb}+ A^{\Gb}_n \d x^n) \nn\\
&=& (g_{mn}+g_{\Ga\Gb}A_m^{\Ga} A_n^{\Gb}) \d x^m \d x^n + 2 g_{\Ga \Gb} A^{\Gb}_m \d x^m \d x^{\Ga} + g_{\Ga \Gb} \d x^{\Ga} \d x^{\Gb} \ ,\label{extg}\\
\tB&=&\frac{1}{2} B_{mn} \d x^m \w \d x^n + g_{\Ga\Gb}\ A^{\Gb}_m \d x^m \w \d x^{\Ga} + \frac{1}{2} B_{\Ga\Gb} \d x^{\Ga} \w \d x^{\Gb} \ ,\label{extB}
\eea
where we introduced $(\d x^{m=1\dots \di}, \d x^{\Ga=1 \dots \dg})$, the one-forms defined locally on this bigger space. Because one considers
 $(\di+\dg) \times (\di+\dg)$ matrices, one could naively consider the T-duality group to be enhanced to $O(\di+\dg,\di+\dg)$. Nevertheless,
the form of the matrices is constrained, and should remain the same under the transformations. In particular, the last
$\dg$ lines of $\tg+\tB$ are totally fixed and should be left unchanged. Therefore, the group is only $O(\di,\di+\dg)$.

As far as we know, this extension of the metric and $B$-field has only been considered as a trick to work out the T-duality transformations in
heterotic string, but not as actual target space fields on a $(\di+\dg)$-dimensional space. Here we will go further and consider a
higher dimensional theory. Provided its fields are given by expressions close to (\ref{extg}) and (\ref{extB}) (which we will call the ``heterotic ansatz''),
 we will show that this theory is equivalent to the (abelian bosonic) heterotic string effective description at order $\ap$. The heterotic T-duality will then be more natural
 in this equivalent higher dimensional theory.\\

 More precisely, we are going to consider a theory on a $\tD$-dimensional target space, where $\tD=D+\dg$, and $D\geq \di$ should correspond eventually to
 the dimension of the standard heterotic target space (usually $D=10$). We take for this $\tD$-dimensional theory the following action
\beq
\tilde{S}=\frac{1}{2\kappa_{\tD}^2} \int \d^{\tD} x \sqrt{|\tg|} e^{-2\tp}
\left[ \tR +4 |\d \tp|^2 -\frac{1}{2} |\tH|^2 + \frac{\ap}{4} \tr (\tR_+^2) \right] \ ,
\eeq
with the metric $\tg$, the dilaton $\tp$, and the two-form $B$-field $\tB$. The $H$-flux is defined as
\beq
\tH=\d \tB + \frac{\ap}{4} {\rm CS}(\to) \ ,
\eeq
where $\to$ is a spin connection involving $\tH$. We refer to the core of the paper for more details. This
action is very close the (bosonic) heterotic string effective description at order $\ap$. The main differences are that it is $\tD$-dimensional,
 and that there are no gauge field.

Then, we derive at order $\ap$ the Bianchi identity (BI) and the equations of motion (e.o.m.) of this theory. The latter
 are of course very close to the heterotic e.o.m. Nevertheless, the derivation of these equations at order $\ap$ is not so straightforward.
It is complicated by the dependence of the connection $\omega_+$ in the metric, the $B$-field, and the gauge potential. Fortunately, the
 variation of the action with respect to this $\omega_+$ can be written as
\bea
\!\!\!\!\!\!\! \frac{2\kappa^2}{\sqrt{|g|}} \frac{\delta S}{\delta \omega^a_{+\ bc}} &=& -\frac{\ap}{4} \Bigg[ (-*)\left(
\theta^c\w \omega^b_{+\ a} \w E_{B\ 0} \right) + 4 \eta^{bl}\eta^{cd}\eta^{fe} \ \nabla_{+,-\ f} \left(e^{-2\phi}  (\d H)_{lade} \right) \\
&& + 2 e^{-2\phi} \eta^{bl}\eta^{cd} \Big( 2 \nabla_{-\ [a} \left( R_{-\ d|l]} +2 \nabla_{-\ l]} \del_{d} \phi \right)
 + H^k_{\ al} \left( R_{-\ dk} +2 \nabla_{-\ k}\del_{d} \phi \right) \Big) \Bigg] + O(\app) \ ,\nn \\
&&\nn\\
\!\!\!\!\!\!\!{\rm with}&& \!\!\!\!\!\!\!\!\!\!\!\!\! R_{-\ be} + 2 \nabla_{-\ e}\del_b \phi \ =\ E_{g\ 0, eb} + \frac{\eta_{eb}}{2} E_{\phi\ 0} - \frac{1}{2} \eta_{bc} \eta_{ed} (-*)\left(
\theta^c\w \theta^d \w e^{2\phi} E_{B\ 0} \right) + O(\ap) \ ,
\eea
where $E_{g\ 0, eb}$, $E_{\phi\ 0}$ and $E_{B\ 0}$ are the Einstein equation, the dilaton and the $B$-field e.o.m. at order $\a$, and we refer to appendix
\ref{apheteom} for the other notations. Since the BI
for $H$ is of order $\ap$, the variation of the action with respect to $\omega_+$ is given by the e.o.m. at order $\a$, up to
 terms of order $O(\app)$. Therefore, provided that the e.o.m. are satisfied order by order in $\ap$, this variation (of order $\ap$) can be consistently discarded. This is the
 result of a lemma worked out in \cite{BR}, that we rederived in details in appendix \ref{apheteom}. This derivation allowed us to verify that this result
 does not depend on the dimension of the space, nor on its signature. This way, we could use it for the derivations of the $\tD$-dimensional e.o.m.

Finally, we use a particular ansatz for the $\tD$-dimensional space and fields decomposed on the $D+\dg$ dimensions. The geometric picture of the
 $\tD$-dimensional space is the following
\begin{center}
\begin{tabular}{cccc}
$U(1)^{\dg}$ & $\hookrightarrow$ & $\tD$-dimensional space(-time) \\
             &                   & $\downarrow$ \\
             &                   & $D$-dimensional space(-time)
\end{tabular}
\end{center}
The $\dg$ circles are fibered over a $D$-dimensional base, via connection one-forms $A^{\Ga}$. The gauge potential $\aaa^{\Ga}$ of heterotic string will end-up being related
 to a connection one-form $A^{\Ga}$, so the gauge fields are incorporated in the geometry. Such a geometric setting for an abelian gauge group is of course
not surprising. Indeed, a gauge theory can be viewed geometrically as a principle bundle (a gauge bundle), where the fiber is the gauge group
 (here $U(1)^{\dg}$). What is less obvious in our construction is to consider a $\tD$-dimensional theory on this geometric setting, and to choose accordingly ans\"atze for
the fields living there. For the metric on this space, it might still be straightforward to choose an ansatz, but it is not the case for the other
 $\tD$-dimensional fields, in particular the $B$-field\footnote{The dependence of the $B$-field on the gauge potential will play a crucial role. See footnote
\ref{footBgauge} of section \ref{seceq}.}. We were inspired here by the pseudo fields of the heterotic T-duality: the
 ``heterotic ansatz'' we take for our $\tD$-dimensional fields is very close\footnote{One difference is that we do not use the Cartan matrix, because we focus here on the
 abelian case.} to the expressions (\ref{extg}) and (\ref{extB}) of the pseudo fields. The heterotic T-duality is then very natural in this higher
dimensional set-up.\\

The main result of this paper is that using this ansatz, we show explicitly that the $\tD$-dimensional action, e.o.m. and BI are equivalent to those of
 the (abelian bosonic) heterotic string effective description, at order $\ap$. To get this result, we need to match some quantities on both sides, like
 the connection one-form and the gauge potential. We paid a particular attention to the $\ap$ dependences and corrections. They play an important role in
 discarding some terms. Finally, note that the BI are indeed the same, but we do not treat here their integrated versions, and the associated global aspects,
 which can be involved. For global aspects of supersymmetric solutions of the type of \cite{GP}, see \cite{FY, EM}.

As a supplement, we also show an equivalence between heterotic string supersymmetry (SUSY) conditions, and some $\tD$-dimensional conditions. To do so, the ``heterotic ansatz'' is refined to such a space
\begin{center}
\begin{tabular}{cccc}
$U(1)^{\dg}$ & $\hookrightarrow$ & $\nnn$          &             \\
             &                   & $\downarrow$    &             \\
             &                   & $\mmm$ & $\times \quad$ $(D-d=4)$ Minkowski \\
             &                   & \multicolumn{2}{c}{$\underbrace{\phantom{....................................................}}\quad$}\\
             &                   & \multicolumn{2}{c}{$D$-dimensional space-time}
\end{tabular}
\end{center}
where the $(d+\dg)$-dimensional manifold $\nnn$ is equipped with appropriate almost hermitian structures. With $\JN$ and $\ON$ respectively the fundamental
 two-form and the maximally $(\frac{d+\dg}{2},0)$-form on $\nnn$, we then propose to consider the following conditions
\bea
&& \d (e^{-2\tp} \ON)=0 \\
&& \d (e^{-2\tp} \JN^{\frac{d+\dg}{2}-1}) =0 \\
&& i(\del-\ov{\del}) \JN= \tilde{H} \ .\label{eq:introSUSY}
\eea
These are $\tD$-dimensional geometric conditions similar to the standard heterotic string SUSY conditions \cite{S, Hu}. Therefore, we name them for convenience the ``$\tD$-dimensional SUSY conditions'', even if we do not consider any SUSY transformation on the $\tD$-dimensional theory (SUSY is rather unlikely to be realised in this theory, as further discussed in section \ref{secSUSY}). We show the equivalence of this set of conditions and the heterotic string SUSY conditions, including in particular the hermitian Yang-Mills condition on the gauge fields.

Note that in this paper, we mean by equivalence the fact that one set of equations can be rewritten into the other, if the ansatz is plugged-in. It is, in a sense, a computational result. In section \ref{secCcl}, we come back to the question of whether one can consider this rewriting as a proper dimensional reduction (this is not what we do here). In particular, some additional modes could a priori be considered in such a reduction, and discarding them could correspond to imposing the chirality of heterotic string. From now on, we only mean by equivalence the result of the rewriting, and leave for future work the possibility of understanding it as a dimensional reduction.\\

Given this equivalence, we study some important consequences for heterotic string solutions. From the $\tD$-dimensional point of view, the $U(1)^{\dg}$
 circles of the ``heterotic ansatz'' are not special. In particular, if the $D$-dimensional space would also contain some circles, one could not distinguish
 them from the others. It is the ``projection'' to heterotic string which forces us to separate the $\tD$-dimensional space into a $\dg$-dimensional part,
which we claim to give the gauge fields, and a $D$-dimensional part which is said to be the real space-time. Therefore, going to heterotic string, one needs
 to distinguish which circle becomes geometric and which one gives a gauge field. As a consequence, by simply exchanging directions in the $\tD$-dimensional
 theory, one can end-up with very different set-ups in heterotic string. Put differently, two different solutions of heterotic string could be
related trivially in the $\tD$-dimensional theory by simply exchanging some directions.

We illustrate this idea with known supersymmetric solutions \cite{DRS, BD, GP, FY, BBFTY}. The first solution has some non-trivial abelian gauge field, but its
 ten-dimensional space-time is given by four-dimensional Minkowski times the K\"ahler manifold $T^2 \times K3$. The second solution on the contrary does not
 have any gauge field, but lives on four-dimensional Minkowski times a non-trivial fibration of $T^2$ over $K3$, which is a non-K\"ahler manifold. These two
solutions have been related in various manners in the literature \cite{BTY, Se, A, EM, AMP, MSp}. As explained here, these two solutions are trivially related in our $\tD$-dimensional
 theory by a simple exchange of two circle directions. We also conjecture the existence of a non-supersymmetric solution which would be non-K\"ahler and have
 non-trivial gauge field. Such a solution would be the same as the others from the $\tD$-dimensional point of view, up to the exchange of only one circle.

Finally, we come back to the heterotic T-duality. Its formulation in terms of the $\tD$-dimensional theory and the ``heterotic ansatz'' is now very natural.
We show that the exchange of directions just mentioned can be encoded in a particular heterotic T-duality, not given by the Buscher rules \cite{B}. We also
 study Buscher T-dualities. In particular, T-dualising along the $A^{\Ga}$ one-form direction could lead to new non-geometric solutions. We also
discuss what happens when T-dualising along the fiber, and compare the behavior of the heterotic solutions with those of type II SUGRA. This leads us
 to some comments on a possible GCG set-up in heterotic string, building-up on a related discussion in \cite{AMP}.\\

The paper is organized as follows. In section \ref{sechetconv}, we give the action of the (bosonic) heterotic string effective description at order $\ap$, its e.o.m.
 and BI. Related conventions on forms and Riemannian geometry are given in appendix \ref{apconv}. The derivation of the heterotic string e.o.m.,
 with particular emphasis on the proof of the lemma of \cite{BR} on $\omega_+$, is given in appendix \ref{apheteom}. In section \ref{seceq}, we first
 introduce the $\tD$-dimensional theory with its action, e.o.m. and BI. Then we detail the ``heterotic ansatz''. Using it, we prove the equivalence
 with the (abelian bosonic) heterotic string effective description at order $\ap$, as far as the action, e.o.m. and BI are concerned. The technical details are given in
 appendix \ref{apreduc}. We end the section with comments on an extension to the non-abelian case. The ``$\tD$-dimensional SUSY conditions'', and their
 equivalence with the heterotic ones, are discussed in section \ref{secSUSY}. In section \ref{secsol}, we detail and illustrate the relations just discussed
between K\"ahler and non-K\"ahler heterotic solutions. We come back to T-duality in great details in section \ref{secTd}, and use it for the various
 applications just mentioned. We conclude this paper with some remarks in section \ref{secCcl}.

Finally, let us mention in a footnote\footnote{Technically, the reduction from $\tD$ to $D$ dimensions of the NSNS action
 at order $\a$ has been done in \cite{MS}. Nevertheless, the purpose of the reduction done there was different; in particular the various components of the
 fields (in particular the off-diagonal component of the $B$-field) were not fully specified as we do, and so the heterotic effective action was not
 recognized after the reduction. In our analysis, not only we specify the fields so that we eventually recognize the heterotic effective action, but we do
 so at order $\ap$.

During the completion of this project appeared the paper \cite{MSp}. There, it is mentioned that the heterotic equations of motion in nine dimensions
at order $\a$ with only one gauge potential are equivalent to the NSNS e.o.m. in ten dimensions, provided one takes for
 the NSNS fields a similar ansatz as the one we took. On this question of the e.o.m., our analysis goes further. Indeed, we consider theories at order
 $\ap$, with several abelian gauge potentials, and we work out the equivalence of the e.o.m. in any dimension.} few similarities with existing papers.

\section{Heterotic string effective description in ten dimensions}\label{sechetconv}

In this section we briefly review the (bosonic) heterotic string effective description in ten dimensions at order $\ap$. The bosonic massless spectrum
 of the ten-dimensional heterotic string is given by the metric $g_{MN}$, the $B$-field $B_{MN}$, the dilaton $\phi$, and
 the Yang-Mills gauge potential $\aaa^{\Ga}_{M}$, where $M,\ N$ stand for space-time indices and $\Ga$ is a color index.

The associated gauge group $G$
 has generators $t_{\Ga}$ in the algebra $\Gg$, with $\Ga=1 \dots {\rm dim}\ \Gg$. $\aaa^{\Ga}=\aaa^{\Ga}_{M} \d x^{M}$ is a space-time
 one-form, where $\{\d x^{M}\}$ is a generic space-time one-form\footnote{See appendix \ref{apforms} for our conventions on forms.} basis, and the full
gauge potential is denoted $\aaa=\aaa^{\Ga} t_{\Ga}$. One can get the field strength for $\aaa$ by acting with the gauge covariant derivative
\beq
\fff^{\Ga} t_{\Ga}= \fff= (\d + \aaa\w) \aaa \ ,\ \fff^{\Ga}= \d \aaa^{\Ga} + \frac{1}{2} f^{\Ga}_{\ \ \Gb \Gc} \aaa^{\Gb} \w \aaa^{\Gc} \ ,\label{defF}
\eeq
where $\d$ is the exterior derivative, and $[t_{\Gb},t_{\Gc}]=f^{\Ga}_{\ \ {\Gb}{\Gc}} t_{\Ga}$ defines the structure constants\footnote{Note that for
 a (semi-) simple Lie algebra $\Gg$ in the adjoint representation, these conventions correspond to choose anti-hermitian generators.}. We introduce the
 trace $\tr$ for the generators. When $G=SO(32)$,
 one should consider the fundamental representation, and so the trace $\tr$ is the corresponding one. No such representation exists for $G=E_8 \times E_8$,
 which has to be rather considered in the adjoint representation. In that case, one can replace $\tr$ by $\frac{1}{30}\tr_{{\rm ad}}$ where $\tr_{{\rm ad}}$
 is the trace in the adjoint representation; indeed for the subgroup $SO(16)\times SO(16)$, these two are equal.

The $B$-field can be written in terms of a two-form $B$. At order $\a$, acting on it with the exterior derivative gives the associated $H$-flux. At order
 $\ap$, the definition of $H$ is corrected as
\beq
H=\d B + \frac{\ap}{4} \left( {\rm CS}(\omega_+) - {\rm CS}(\aaa)\right) + O(\app) \ ,\label{defHhet}
\eeq
where $\omega^a_{+\ bM}=\omega^a_{\ bM}+ \frac{1}{2} H^a_{\ bM}$ is the $M$-coefficient of the connection\footnote{See appendices \ref{apcovder} and
 \ref{aphetcon} for our conventions on covariant derivatives and connections.} one-form $\omega^a_{+\ b}$ (\ref{connecpm}). The local frame indices are
denoted $a,b, \dots$, and $\omega^a_{\ b}$ is the connection one-form associated to Levi-Civita. The curvature two-forms associated to $\omega^a_{\ b}$,
respectively $\omega^a_{+\ b}$, are denoted $R^a_{\ b}$, respectively $R^a_{+\ b}$. The curvature two-forms can be obtained as in (\ref{Riem}). The two
 ${\rm CS}$ in (\ref{defHhet}) denote the Chern-Simons terms:
\beq
{\rm CS}(\aaa)=\tr(\fff\w \aaa - \frac{1}{3} \aaa\w \aaa \w \aaa) \ ,\
{\rm CS}(\omega_+)=-\left(R^a_{+\ b}\w \omega^b_{+\ a} - \frac{1}{3} \omega^a_{+\ b} \w \omega^b_{+\ c}\w \omega^c_{+\ a} \right) \ .
\eeq
As a convention, the trace on the local frame indices will always bring a minus sign as in the definition of ${\rm CS}(\omega_+)$. Acting with
the exterior derivative on the ${\rm CS}$ terms, one gets, using the cyclicity of the trace,
\beq
\d {\rm CS}(\aaa) = \tr(\fff \w \fff) \ , \ \d {\rm CS}(\omega_+) = \tr(R_+ \w R_+) = -  R^a_{+\ b} \w R^b_{+\ a} \ .
\eeq
Therefore, if $\d B$ is globally defined, the Bianchi identity (BI) of the $H$-flux is
\beq
\d H= \frac{\ap}{4} \left( \tr(R_+ \w R_+) - \tr(\fff \w \fff)\right) + O(\app) \ . \label{BI}
\eeq
In presence of an $NS5$-brane, the BI gets an additional source term (see for instance \cite{HLMM} for a derivation). Note that this
term is also of order $\ap$, and so is always $\d H$. From now on, we will not consider any $NS5$-brane.\\

For two $k$-forms $A$ and $B$ in a $D$-dimensional space-time we introduce the notation
\beq
A \cdot B = B \cdot A = \frac{A_{m_1 \dots m_k} B^{m_1 \dots m_k}}{k!} \ \ ,\ \ A \w *_D B = B \w *_D A = \d^{D} x\  \sqrt{|g|} \ A \cdot B \ , \label{FormAB}
\eeq
where $*_D$ denotes the Hodge star (\ref{Hodge*}) in the $D$-dimensional space-time, and $|g|$ is (the absolute value of) the
 determinant of the metric. In addition, for $A=B$, we denote $A \cdot A = |A|^2$. Then, the bosonic part of the heterotic string effective
action at order $\ap$ is given by
\beq
S=\frac{1}{2\kappa^2} \int \d^{10} x \sqrt{|g|} e^{-2\phi}
\left[ R +4 |\d \phi|^2 -\frac{1}{2} |H|^2 + \frac{\ap}{4} (\tr (R_+^2)-\tr (\fff^2)) + O(\app) \right] \ ,\label{action}
\eeq
where $2\kappa^2=(2\pi)^7 (\alpha^\prime)^4$, $\alpha^\prime=l_s^2$ , and
\beq
\tr (\fff^2)= \tr(t_{\Ga}t_{\Gb})\ \fff^{\Ga}\cdot \fff^{\Gb}\ \ ,\ \ \tr (R_+^2)= - R^a_{+\ b} \cdot R^b_{+\ a} = \frac{1}{2}R_{+\ abMN} R_+^{abMN} \ .\label{trR2}
\eeq
The conventional minus sign of the last trace is taken so that it disappears in the last equality, thanks to the antisymmetry property\footnote{This property
is discussed in appendix \ref{aphetcon}. Note also that one could take all indices to be either space-time or local frame in the
 last term of (\ref{trR2}), so that the Kretschmann scalar with respect to $\omega_+$ appears.} of the Riemann tensor.

Deriving equations of motion (e.o.m.) at order $\ap$ out of this action turns out to be rather involved. The derivation is complicated by the
fact $\omega_+$ depends on the metric, the $B$-field, and the gauge potential. Fortunately, it turns that the variation of
the action with respect to $\omega_+$, which is of order $\ap$, is related to the e.o.m. at order $\a$. Indeed, one has
\bea
\!\!\!\!\!\!\! \frac{2\kappa^2}{\sqrt{|g|}} \frac{\delta S}{\delta \omega^a_{+\ bc}} &=& -\frac{\ap}{4} \Bigg[ (-*)\left(
\theta^c\w \omega^b_{+\ a} \w E_{B\ 0} \right) + 4 \eta^{bl}\eta^{cd}\eta^{fe} \ \nabla_{+,-\ f} \left(e^{-2\phi}  (\d H)_{lade} \right) \label{dSdo}\\
&& + 2 e^{-2\phi} \eta^{bl}\eta^{cd} \Big( 2 \nabla_{-\ [a} \left( R_{-\ d|l]} +2 \nabla_{-\ l]} \del_{d} \phi \right)
 + H^k_{\ al} \left( R_{-\ dk} +2 \nabla_{-\ k}\del_{d} \phi \right) \Big) \Bigg] + O(\app) \ ,\nn \\
&&\nn\\
\!\!\!\!\!\!\!{\rm with}&& \!\!\!\!\!\!\!\!\!\!\!\!\! R_{-\ be} + 2 \nabla_{-\ e}\del_b \phi \ =\ E_{g\ 0, eb} + \frac{\eta_{eb}}{2} E_{\phi\ 0} - \frac{1}{2} \eta_{bc} \eta_{ed} (-*)\left(
\theta^c\w \theta^d \w e^{2\phi} E_{B\ 0} \right) + O(\ap) \ ,
\eea
where $E_{g\ 0, eb}$, $E_{\phi\ 0}$ and $E_{B\ 0}$ are the Einstein equation, the dilaton and the $B$-field e.o.m. at order $\a$. Since the BI
for $H$ is of order $\ap$, the variation of the action with respect to $\omega_+$ is indeed given by the e.o.m. at order $\a$, up to
 terms of order $O(\app)$. Therefore, provided these e.o.m. are satisfied, one can discard the variation with respect to $\omega_+$.
 This is the result of a lemma proven in \cite{BR}, that we rederive in details in appendix \ref{apheteom} (see in particular (\ref{dSdoap})). We also derive
 in this appendix the whole set of e.o.m. at order $\ap$. The result is the following: provided the e.o.m. are satisfied order by order in $\ap$,
 they can be written as
\bea
R -\frac{1}{2} |H|^2 + 4(\nabla^2 \phi - |\d \phi|^2 ) + \frac{\ap}{4} (\tr (R_+^2)-\tr (\fff^2)) &=& 0 + O(\app) \\
\!\!\!\!\!\!\!\!\!\!\!\!\!\!\!\!\!\!\!\!\! R_{MN} - \frac{1}{2}\ \iota_{M} H \cdot \iota_{N} H + 2\nabla_{M}\nabla_{N} \phi
+\frac{\ap}{4} (\tr (\iota_{M} R_+ \cdot \iota_{N} R_+)-\tr (\iota_{M} \fff \cdot \iota_{N} \fff)) &=& 0 + O(\app) \\
\d(e^{-2\phi} * H) &=& 0 + O(\app) \\
e^{2\phi} \d(e^{-2\phi} * \fff) +  \aaa \w * \fff - * \fff \w \aaa - \fff \w *H &=& 0 + O(\ap) \ ,\label{hetFeom}
\eea
where in the last equation one should only consider the order $\a$ in $H$. We introduced for a $k$-form $A$ the notation
$\iota_M A = \iota_{\d x^M} A$ out of (\ref{contraction}).

\section{Equivalence with a higher dimensional theory}\label{seceq}

In this section, we first introduce a $\tD$-dimensional bosonic theory, and derive its equations of motion and Bianchi identity. Then, choosing some
particular ansatz for the $\tD$-dimensional space, and accordingly for the fields of this theory, we show that this action, e.o.m. and BI, are equivalent
to the abelian heterotic string effective action (\ref{action}) (considered in a space of arbitrary dimension), e.o.m., and BI, at order $\ap$. We end the
 section with a few comments on a generalization to the non-abelian case.

We recall from the comments below (\ref{eq:introSUSY}) that we do not mean by equivalence performing a proper dimensional reduction, but rather obtaining various known equations by rewriting others, using an ansatz.

\subsection{A higher dimensional theory}

We consider the following action in a $\tD$-dimensional space (the properties of this space, in particular its signature, or the parity of
$\tD$, do not need to be specified in this section)
\beq
\tilde{S}=\frac{1}{2\kappa_{\tD}^2} \int \d^{\tD} x \sqrt{|\tg|} e^{-2\tp}
\left[ \tR +4 |\d \tp|^2 -\frac{1}{2} |\tH|^2 + \frac{\ap}{4} \tr (\tR_+^2) \right] \ , \label{StD}
\eeq
where $\kappa_{\tD}$ is a constant with the appropriate dimensionality, that is going to be fixed. Similarly to the heterotic string, the fields to be
considered in this theory are the metric $\tg$, the dilaton $\tp$, and the two-form $B$-field $\tB$. The $H$-flux is defined as
\beq
\tH=\d \tB + \frac{\ap}{4} {\rm CS}(\to) \ ,\label{Hgene}
\eeq
where $\to$ is defined with $\tilde{\omega}^{\ta}_{+\ \tb\tM}=\tilde{\omega}^{\ta}_{\ \tb\tM}+\frac{1}{2} \tH^{\ta}_{\ \tb\tM}$,
and $\tM$, respectively $\ta$, denote $\tD$-dimensional space, respectively local frame, indices. $\tilde{\omega}^{\ta}_{\ \tb\tM} \d x^{\tM}$ is
the standard connection one-form associated to Levi-Civita. Associated curvature two-forms $\tR^{\ta}_{\ \tb}$ and
$\tR^{\ta}_{+\ \tb}$ are also defined similarly to the heterotic string, and so is the $\tr (\tR_+^2)$. Provided $\d \tB$
is globally defined, we get the following Bianchi identity for $\tH$,
\beq
\d \tH = \frac{\ap}{4} \tr(\tR_+ \w \tR_+) \ . \label{tBI}
\eeq

This action and the fields considered are clearly similar to the heterotic string effective description at order $\ap$, discussed in the previous section,
with the differences that the dimension is here $\tD$, and that there is no gauge field. As a consequence, one can follow the same procedure
to derive the e.o.m. from this action. As emphasized in appendix \ref{apheteom}, the derivation of these e.o.m. at order
$\ap$, even if rather involved, does not depend on the properties of the space, in particular not on its dimension, nor on
the signature of this space. Therefore, one finally obtains the same result, simply without any gauge field:
\bea
\tR -\frac{1}{2}|\tH|^2 + 4(\tna^2 \tp - |\d \tp|^2 ) + \frac{\ap}{4} \tr (\tR_+^2) &=& 0 \label{dileom}\\
\tR_{\tM\tN} - \frac{1}{2}\ \iota_{\tM} \tH \cdot \iota_{\tN} \tH  + 2\tna_{\tM}\tna_{\tN} \tp
+ \frac{\ap}{4} \tr (\iota_{\tM} \tR_+ \cdot \iota_{\tN} \tR_+) &=& 0 + O(\app) \label{Einstein}\\
\d(e^{-2\tp} *_{\tD} \tH) &=& 0 + O(\app) \ . \label{Beom}
\eea
Given we do not consider terms of order $O(\app)$ in the action (\ref{StD}), the derived dilaton e.o.m. (\ref{dileom}) is valid at all orders. On the
contrary, for the other two equations, we have to follow the same reasoning as for the heterotic string e.o.m. which involves the variation with respect
 to $\to$. This leads to corrections at order $\app$.

\subsection{The ``heterotic ansatz''}\label{sechetans}

We are now going to take an ansatz for our $\tD$-dimensional space, and accordingly for the fields of the theory. The $\tD$-dimensional action, e.o.m.
 and BI, will then simply reduce to those of the abelian heterotic string effective description at order $\ap$, provided we match some quantities.
 As discussed in the Introduction, this ansatz is motivated by the heterotic T-duality.\\

The ansatz consists first in splitting the $\tD$-dimensional space into two parts of dimensions denoted $D$ and $\dg$: $\tD=D+\dg$.
 The $D$-dimensional part will correspond eventually to the ten-dimensional space-time
 of heterotic string ($D=10$). But for sake of generality, we keep a generic $D$ all along, and do not specify the properties of this space. The
 $\dg$-dimensional part will correspond in the end to the fiber of the gauge bundle, hence the $g$ index. Similarly, we do not specify the parity of $\dg$ or the
signature of this space.

Nevertheless, the $\dg$-dimensional space is further restricted: we take it to be
 $U(1)^{\dg}$, i.e. it is made of $\dg$ commuting circles, fibered other the $D$-dimensional part.
\begin{center}
\begin{tabular}{cccc}
$U(1)^{\dg}$ & $\hookrightarrow$ & $\tD$-dimensional space(-time) \\
             &                   & $\downarrow$ \\
             &                   & $D$-dimensional space(-time)
\end{tabular}
\end{center}
The directions of the circles are locally given by the real one-forms $\d x^\Ga$; we denote the
indices of this space with $\Ga$, since it will correspond eventually to the fiber of the gauge bundle. The fibrations are encoded in connections $A^{\Ga}$,
 which are locally defined one-forms, living on the $D$-dimensional space. $\d x^{\Ga} +A^{\Ga}$ are then the well-defined one-forms on the total
 $\tD$-dimensional space. We label the $D$-dimensional indices as $M,N$. If we take for this part of the space a generic metric and basis of
 one-forms so that $\d s_{D}^2=g_{MN} \d x^M \d x^N$, then the total metric is given by
\beq
\d s_{\tD}^2=g_{MN} \d x^M \d x^N + g_{\Ga \Gb} (\d x^{\Ga}+ A^{\Ga})(\d x^{\Gb}+ A^{\Gb}) \ . \label{hetansmet}
\eeq
As a further choice in this ansatz, we will consider from now on the metric $g_{\Ga \Gb}$ to be constant. As we will see eventually, this is a good choice to recover the abelian heterotic string effective description at order $\ap$.

For later convenience, we introduce the two following basis of one-forms on the $\tD$-dimensional space-time
\bea
&& \bbb_1= (\{\d x^M \}, \{\d x^{\Ga} \}) \ ,\nn\\
&& \bbb_2= (\{\d x^M \}, \{\d x^{\Ga}+ A^{\Ga} \}) \ .
\eea
The basis $\bbb_2$ is, up to constant linear transformations, the vielbein basis (local frame), at least as far as the fiber $U(1)^{\dg}$ is concerned.
 The metric is block-diagonal in this basis, and Hodge star
computations are then simpler. On the contrary, computations of the Levi-Civita connection need to be done in the basis $\bbb_1$, which is a coordinate basis.

Let us now give our ansatz for the other fields on this $\tD$-dimensional space. No field will depend on
 the $U(1)^{\dg}$ coordinates. In particular, the dilaton $\tp$ is taken to depend only on the $D$-dimensional coordinates,
so for simplicity we identify it with the $D$-dimensional dilaton up to a constant $\varphi$: $\tp=\phi+\varphi$. Finally,
 the $B$-field has various
components on the different parts of the $\tD$-dimensional space. We express it in the basis $\bbb_1$ as
\beq
\tB=B + B_g  + c g_{\Ga\Gb}\ A^{\Ga} \w \d x^{\Gb}\ , \label{hetansB}
\eeq
where $B$ is the $D$-dimensional $B$-field, $B_g$ is a closed two-form with components along the fiber directions
$\d x^{\Ga}$ only, and the last term is a mixed base-fiber component. $c$ is a constant to be fixed.

Note that this ansatz for the metric and the $B$-field does correspond to the pseudo fields (\ref{extg}) and (\ref{extB}) proposed to be considered for
 heterotic T-duality. We will come back to this relation in section \ref{secTd}.\\

Since no field depends on the circles coordinates $x^{\Ga}$, the exterior derivative $\d$ on the
$\tD$-dimensional space can be written in the same way as the $D$-dimensional one. So the $\tD$-dimensional $H$-flux, defined as in (\ref{Hgene}), becomes
\beq
\tH=\d B + c g_{\Ga\Gb}\ \d A^{\Ga} \w \d x^{\Gb} +    \frac{\ap}{4} {\rm CS}(\to)  \ .
\eeq
Let us define the following forms
\bea
&& H= \d B - c g_{\Ga\Gb}\ \d A^{\Ga} \w A^{\Gb} + \frac{\ap}{4} {\rm CS}(\omega_+)  \ ,\label{defHanshet}\\
&& F^{\Ga}=\d A^{\Ga} \ ,
\eea
where $\omega_+$ is the connection associated to the $D$-dimensional space alone, with metric $\d s_{D}^2$, and the $H$ entering its definition being the
 form just defined. The form $H$ introduced will then eventually correspond to the heterotic string $H$-flux at order $\ap$. Then we can rewrite
the $\tD$-dimensional $H$-flux as
\beq
\tH=H + c g_{\Ga\Gb}\ F^{\Ga} \w (\d x^{\Gb}+ A^{\Gb}) +\frac{\ap}{4} \left({\rm CS}(\to) - {\rm CS}(\omega_+) \right) \ .
\eeq

\subsection{The equivalence}

We are now going to plug the ansatz just discussed into the $\tD$-dimensional theory, and show the equivalence with the abelian heterotic string effective
 description at order $\ap$ (see comments below (\ref{eq:introSUSY})). But we need at first to carefully focus on the $\ap$ dependence, and discuss what terms can be discarded at order $\ap$. Then, we present the main results of the rewriting of the action, the e.o.m. and the BI, once the ``heterotic ansatz'' is taken into account. We leave the technical details to appendix \ref{apreduc}. Finally, we match the quantities of both theories to show the equivalence.

\subsubsection{The $\ap$ dependence}\label{secap}

In order to recover the heterotic string effective description at order $\ap$, we have to discuss the dependence of our ansatz in $\ap$. A crucial point
is the relation between the connection $A^{\Ga}$ and the heterotic gauge potential $\aaa^{\Ga}$, which will be established precisely in section
 \ref{secmatch}: it will involve some $\ap$ dependence. The reason is dimension wise. As one can figure out from the heterotic string effective action,
the gauge potential $\aaa^{\Ga}_M$ has the dimension of an inverse length, provided the generators $t_{\Ga}$ are dimensionless. On the other hand,
 the connection one-form $A^{\Ga}$ has the dimension of a length, so $A^{\Ga}_M$ is dimensionless. Therefore, up to dimensionless coefficients, the two
objects will be identified with an $\sqrt{\ap}$ factor: $A^{\Ga}_M \sim \sqrt{\ap} \aaa^{\Ga}_M$. Let us discuss the consequences.

We first consider the Riemann tensor $\tR^{\tM}_{\ \ \tN \tP \tQ}$ in the basis $\bbb_1$. It is given only in terms of connection coefficients
 $\tG^{\tM}_{\tN\tP}$ and derivatives
 of them. As computed in appendix \ref{apreduc}, these coefficients only differ from the purely $D$-dimensional connection coefficients (those related to $g_{MN}$)
 by terms depending on $A^{\Ga}_M$. So $\tR^{\tM}_{\ \ \tN \tP \tQ}$ only differs from its purely $D$-dimensional counterpart by dependences on $A^{\Ga}_M$. The same goes for the spin connection
 $\tilde{\omega}^{\ta}_{\ \tb}$ with respect to $\omega^{a}_{\ b}$. Indeed, one can compare $\tG^{\ta}_{\tb\tc}$ and $\Gamma^{a}_{bc}$ by looking at the
definition\footnote{The first term in (\ref{gabc}) is related to the derivative of a vielbein. The metric $g_{\Ga\Gb}$ being constant, the only non-trivial
terms obtained from the derivative of this vielbein are either given by the derivative of
the purely $D$-dimensional vielbein or by terms depending on $A^{\Ga}_M$. The study of the second term in (\ref{gabc}) is more involved: it is given by
 $\tG^{\tM}_{\tN \tP} \te_{\tc}^{\ \tP}$. The term $\tG^{\tM}_{\tN P} \te_{\tc}^{\ P}$ only differs from its $D$-dimensional counterpart by terms depending on
$A^{\Ga}_M$, because both $\tG^{\tM}_{\tN P}$ and $\te_{\tc}^{\ P}$ do. For $\tP=\Ga$, it turns out that $\tG^{\tM}_{\tN \Ga}$ is directly dependent on $A^{\Ga}_M$
 (see appendix \ref{apreduc}). So the second term in (\ref{gabc}) also differs from its $D$-dimensional counterpart by $A^{\Ga}_M$ dependent terms, and
 we conclude that the same goes for $\tG^{\ta}_{\tb\tc}$ with respect to $\Gamma^{a}_{bc}$.} (\ref{gabc}).
 Let us now consider $\tilde{\omega}^{\ta}_{+\ \tb}$: the difference with $\tilde{\omega}^{\ta}_{\ \tb}$ is given by $\tH$, which differs from $H$ either by
 a $A^{\Ga}_M$ dependent term, or by an $\ap$ order term. So $\tilde{\omega}^{\ta}_{+\ \tb}$ also differs from $\omega^{a}_{+\ b}$ by terms depending on
$A^{\Ga}_M$ or by terms of order $\ap$. The same goes for the Riemann tensor $\tR^{\tM}_{+\ \tN\tP\tQ}$ with respect to $R^{M}_{+\ NPQ}$, as can be seen
either from (\ref{Riem}) or from (\ref{Riemeps}). We conclude that with the identification $A^{\Ga}_M \sim \sqrt{\ap} \aaa^{\Ga}_M$, we will get that
\bea
&& \frac{\ap}{4} \left({\rm CS}(\to) - {\rm CS}(\omega_+) \right)= 0+ O(\appd)\ ,\label{apo+}\\
&& \frac{\ap}{4} \tr (\tR_+^2)=\frac{\ap}{4} \tr (R_+^2) + O(\appd)\ .\label{apR+}
\eea
So from now on, we will consider
\beq
\tH = H + c g_{\Ga\Gb}\ F^{\Ga} \w (\d x^{\Gb}+ A^{\Gb}) + O(\appd) \ .\label{Hans}
\eeq

Consistently with (\ref{apR+}), one can show in the basis $\bbb_1$ that $\iota_{\Ga} \tR_+= O(\apd)$. Indeed, such a contraction involves either a connection
coefficient with an index $\Ga$,  or an off-diagonal component of the metric $\tg$ (the derivatives with respect to $x^{\Ga}$ do not contribute since they
 are zero). According to appendix \ref{apreduc}, both objects depend on $A^{\Ga}_M$, so we deduce the result. Therefore, we will get in the Einstein equation (\ref{Einstein})
\bea
\frac{\ap}{4} \tr (\iota_{\tM} \tR_+ \cdot \iota_{\tN} \tR_+)&=& \frac{\ap}{4} \delta_{\tM}^M \delta_{\tN}^N \tr (\iota_M \tR_+ \cdot \iota_{N} \tR_+)
+ O(\appd) \nn\\
&=& \frac{\ap}{4} \delta_{\tM}^M \delta_{\tN}^N \tr (\iota_M R_+ \cdot \iota_{N} R_+) + O(\appd) \ . \label{Einstap}
\eea

Let us now use these results to rewrite the $\tD$-dimensional action, e.o.m. and BI, at order $\ap$.

\subsubsection{Let's play}

We rewrite the action, the e.o.m. and the BI of the $\tD$-dimensional theory at order $\ap$, taking into account the ``heterotic ansatz'', and the discussion on the $\ap$
dependence. The computations are detailed in appendix \ref{apreduc}. Note that the Levi-Civita connection has been used both in $\tD$ and $D$ dimensions.
 An important result of these computations is the following:
\beq
\tR=R - \frac{1}{4} g_{\Ga \Gb} F^{\Ga}_{MN} F^{\Gb\ MN} \ , \ |\tH|^2=|H|^2 + \frac{c^2}{2} g_{\Ga \Gb} F^{\Ga}_{MN} F^{\Gb\ MN} + O(\appd) \ .\label{resultRH}
\eeq
Let us now go through the various rewriting.\\

\begin{itemize}
 \item {\bf \underline{Action and dilaton e.o.m.}}\\

If we denote by $|g_g|$ the absolute value of the determinant of the $U(1)^{\dg}$ metric, we can fix our constant $\kappa_{\tD}$ as in a standard dimensional
 reduction by choosing
\beq
\frac{1}{2\kappa_{\tD}^2} \int \d^{\dg} x \sqrt{|g_g|} e^{-2\varphi} = \frac{1}{2\kappa_D^2} \ ,
\eeq
with $\kappa_{10}=\kappa$. Since no field depends on the $U(1)^{\dg}$ coordinates, the action (\ref{StD}) becomes
\beq
\tilde{S}=\frac{1}{2\kappa_D^2} \int \d^{D} x \sqrt{|g_{D}|} e^{-2\phi}
\left[ R +4 |\d \phi|^2 -\frac{1}{2} |H|^2 - \frac{c^2+1}{4} g_{\Ga \Gb} F^{\Ga}_{MN} F^{\Gb\ MN} + \frac{\ap}{4} \tr (R_+^2) + O(\appd) \right] \ ,
\eeq
where we used (\ref{apR+}) and (\ref{resultRH}). In addition, using (\ref{relchris}), one can verify that $\tna^2 \tp=\nabla^2 \phi$, so the dilaton
 e.o.m. (\ref{dileom}) is equivalent to
\beq
R -\frac{1}{2}|H|^2 + 4(\nabla^2 \phi - |\d \phi|^2 ) - \frac{c^2+1}{4} g_{\Ga \Gb} F^{\Ga}_{MN} F^{\Gb\ MN} + \frac{\ap}{4} \tr (R_+^2) = 0 + O(\appd) \ .
\eeq

 \item {\bf \underline{Einstein equation}}\\

As discussed in appendix \ref{apreduc}, the Einstein equation in $\tD$ dimensions (\ref{Einstein}) is more easily decomposed in the basis $\bbb_2$, even if the
 quantities involved are first computed in $\bbb_1$. Eventually, the equation (\ref{Einstein}) is equivalent to the three following equations
\bea
&& R_{MN} - \frac{1}{2} \iota_{M} H \cdot \iota_{N} H  + 2 \nabla_M \del_N \phi \nn\\
&& - \frac{c^2+1}{2} g_{\Ga \Gb} \iota_{M} F^{\Ga} \cdot \iota_{N} F^{\Gb} + \frac{\ap}{4} \tr (\iota_{M} R_+ \cdot \iota_{N} R_+) = 0 + O(\appd) \\
&& \d(e^{-2\phi} *_D F^{\Gb}) - c e^{-2\phi} F^{\Gb}\w *_D H = 0 + O(\appd) \label{FR}\\
&& \frac{1-c^2}{4} g_{\Ga \Gc} g_{\Gb \Gd} F^{\Gc}_{MN} F^{\Gd\ MN}  = 0 + O(\appd)\ , \label{FF}
\eea
where the two diagonal blocks give the first and last equations, and the off-diagonal one gives the second equation.\\

 \item {\bf \underline{$B$-field e.o.m. and $H$ Bianchi identity}}\\

As shown in appendix \ref{apreduc}, the $B$-field e.o.m. in $\tD$ dimensions (\ref{Beom}) is equivalent to the following two equations
\bea
&& \d(e^{-2\phi} *_{D} H)= 0 + O(\appd) \\
&&  c \d\left(e^{-2\phi} *_{D} F^{\Ga} \right) - e^{-2\phi} F^{\Ga} \w  *_{D} H =0 + O(\appd) \ . \label{FH}
\eea

In addition, the Bianchi identity for $\tH$ in $\tD$ dimensions (\ref{tBI}) is equivalent\footnote{Note also that the hypothesis of having $\d \tB$ globally defined is equivalent to having $\d B$ globally
 defined in view of (\ref{hetansB}), since it is implicit that $F^{\Ga}$ should be well-defined.} to
\beq
\d H= \frac{\ap}{4} \tr(R_+ \w R_+) - c g_{\Ga\Gb} F^{\Ga}\w F^{\Gb} + O(\appd) \ ,
\eeq
where we used (\ref{apo+}) and (\ref{Hans}).

\end{itemize}

\subsubsection{Final matching}\label{secmatch}

The action, e.o.m. and BI, just rewritten at order $\ap$, can correspond to the heterotic ones, provided we match some quantities. We first focus on the free
constant $c$. Compatibility\footnote{\label{footBgauge}It is surprising to obtain eventually twice the gauge potential e.o.m., once via the metric and once via the $B$-field.
The dependence of the metric in the gauge potential being rather natural from the geometric construction, one may wonder whether it is really necessary to have
the $B$-field depending on it as well. Nevertheless, this dependence turns out to be crucial at several places. For instance, it is justified by the heterotic
T-duality, as discussed in the Introduction. It also turns out to be important when recovering the SUSY conditions in section \ref{secSUSY}, when relating
 the K\"ahler and non-K\"ahler solutions in section \ref{secsol}, or when applying the Buscher T-duality on them in section \ref{secBusch}. This dependence
 is therefore a non-trivial aspect of the ``heterotic ansatz''.} in full generality of (\ref{FR}) and (\ref{FH}) imposes to fix $c^2=1$. In addition, comparing these equations with the abelian
version of the gauge potential e.o.m. of heterotic string (\ref{hetFeom}), leads us to choose $c=1$. The comparison of the action, or of the flux $H$ with
 the heterotic ones, leads to the same result. Therefore, (\ref{FF}) is trivially satisfied, and we are left with
\beq
\tilde{S}=\frac{1}{2\kappa_D^2} \int \d^{D} x \sqrt{|g_{D}|} e^{-2\phi}
\left[ R +4 |\d \phi|^2 -\frac{1}{2} |H|^2 + \frac{\ap}{4} \tr (R_+^2) - g_{\Ga \Gb} F^{\Ga} \cdot F^{\Gb} + O(\appd) \right] \ ,
\eeq
with the $H$-flux defined as
\beq
H= \d B + \frac{\ap}{4} {\rm CS}(\omega_+) - g_{\Ga\Gb}\ F^{\Ga} \w A^{\Gb} \ ,
\eeq
the Bianchi identity given by
\beq
\d H= \frac{\ap}{4} \tr(R_+ \w R_+) - g_{\Ga\Gb} F^{\Ga}\w F^{\Gb} + O(\appd) \ ,
\eeq
and the following set of equations
\bea
\!\!\!\!\!\!\!\!\!\!\!\!\!\!\!\!\!\!\!\!\!\!\!\!\!\!\!\!\!\!\!\! R -\frac{1}{2}|H|^2 + 4(\nabla^2 \phi - |\d \phi|^2 )
+ \frac{\ap}{4} \tr (R_+^2) - g_{\Ga \Gb} F^{\Ga} \cdot F^{\Gb} &=& 0 + O(\appd) \\
\!\!\!\!\!\!\!\!\!\!\!\!\!\!\!\!\!\!\!\!\!\!\!\!\!\!\!\!\!\!\!\! R_{MN} - \frac{1}{2} \iota_{M} H \cdot \iota_{N} H  + 2 \nabla_M \del_N \phi
+ \frac{\ap}{4} \tr (\iota_{M} R_+ \cdot \iota_{N} R_+) - g_{\Ga \Gb} \iota_{M} F^{\Ga} \cdot \iota_{N} F^{\Gb} &=& 0 + O(\appd) \\
\!\!\!\!\!\!\!\!\!\!\!\!\!\!\!\!\!\!\!\!\!\!\!\!\!\!\!\!\!\!\!\! \d(e^{-2\phi} *_{D} H) &=& 0 + O(\appd)\\
\!\!\!\!\!\!\!\!\!\!\!\!\!\!\!\!\!\!\!\!\!\!\!\!\!\!\!\!\!\!\!\! e^{2\phi} \d(e^{-2\phi} *_D F^{\Ga}) - F^{\Ga}\w *_D H &=& 0 + O(\appd) \ .
\eea
We conclude that the theory defined in $\tD$ dimensions, together with the ``heterotic ansatz'' for its space and fields, is equivalent at order
 $\ap$ (as far as the action, the equations of motion and the Bianchi identity are concerned), to the heterotic string effective description at order $\ap$,
 in ten dimensions, with abelian gauge group, provided we take $D=10$, $c=1$ and fix
\beq
A^{\Ga} = \sqrt{\alpha'} \gamma \aaa^{\Ga} \ , \ g_{\Ga\Gb}=\frac{1}{4	\gamma^2} \tr(t_{\Ga}t_{\Gb}) \ ,\label{matchmet}
\eeq
for $\gamma$ any real constant. Note that such a $g_{\Ga\Gb}$ is constant, which is consistent with our ansatz; its signature did not matter for our purposes, nevertheless we discuss
it in the next section.

We recall from section \ref{secap} that the scaling in $\sqrt{\alpha'}$ of the connection $A^{\Ga}$ is in any case needed
for dimensional reasons. In addition, note that the order of the correction terms does not match exactly for all the equations but the gauge flux one: we get
corrections in $O(\appd)$ while heterotic string effective theory has only $O(\app)$ terms. A more precise check of these terms in the $\tD$-dimensional
 theory could maybe lead to the same order. For our purposes, it does not matter since the theories are only said to match at order $\ap$.

\subsection{Comments on the non-abelian case}\label{secnonab}

In practice, the difference with the non-abelian case can be found in the Chern-Simons term in $H$, and in the gauge potential equation of
 motion: here we only recover the abelian versions of those. Would it be possible to extend the equivalence worked out so far to the non-abelian case?
 Let us point out a few difficulties of such an attempt.\\

From the geometric point of view (and consequently for the metric ansatz), these is no major difficulty in an extension to the non-abelian case. As
discussed in the Introduction, the $U(1)^{\dg}$ piece added to the space-time corresponds to the fiber of the gauge bundle. The
non-abelian extension might then simply correspond to the non-abelian gauge group (viewed as a manifold), with the total space being a
 principle bundle. The non-trivial part in the ``heterotic ansatz'' was rather to consider a non-trivial component of the $B$-field along the gauge
 directions. In addition, this component is somehow related to the Chern-Simons term. The non-abelian generalization of this ansatz is not obvious. In particular, the non-abelian Chern-Simons term can not be obtained by the action of an exterior derivative, while it is the case for the abelian one. Note also that the $A^3$ term in the non-abelian Chern-Simons term would a priori be of order $\appd$ in our procedure. It should therefore be compared with the terms $O(\appd)$ which were discarded.\\

In addition to this question of the $B$-field ansatz, let us discuss another difficulty related to $g_{\Ga\Gb}$ and $\tr(t_{\Ga}t_{\Gb})$. For
 compact (semi-) simple Lie groups, in particular $SO(32)$ or $E_8 \times E_8$, one can diagonalize and rescale
$\tr(t_{\Ga}t_{\Gb})$ so that
\beq
\tr(t_{\Ga}t_{\Gb})=\lambda \ \delta_{\Ga\Gb} \ .
\eeq
Since we used anti-hermitian generators, one gets that $\lambda<0$. In other words, for such groups, the metric $g_{\Ga\Gb}$ we took (\ref{matchmet}) would be negative-definite.
 However, we only focused on the abelian case, i.e. the gauge group was restricted to $G=U(1)^{\dg}$. This is not a simple Lie group, so the result for
$\tr(t_{\Ga}t_{\Gb})$ does not hold for us. Furthermore, the distinction between hermitian and anti-hermitian generators cannot be made at the level of
 the Lie bracket, the covariant derivative, or the Chern-Simons term, in the absence of non-abelian terms in these expressions. On the contrary, one can
choose in the abelian case the desired convention for the generators. Of particular interest for the supersymmetric case that follows, it is possible to
 choose $\tr(t_{\Ga}t_{\Gb})$ to be positive-definite\footnote{Note that this possibility is physically consistent. Indeed, on the one hand, the
 supersymmetric case that follows considers a compact manifold, on which the gauge fields live. In addition, one needs to have an Euclidean signature
 for $g_{\Ga\Gb}$, in order to consider an hermitian metric on the gauge bundle. As explained, having $g_{\Ga\Gb}$ positive-definite is possible in the case of $U(1)^{\dg}$. On the other hand, it is known
 that an effective description of heterotic string on a ten-dimensional space-time which is not Minkowski, but rather includes a compact space, usually leads
 to consider for the gauge group a subgroup of $SO(32)$ or $E_8 \times E_8$. The reason is that only a few modes of the full gauge group remain massless. Then,
$U(1)^{\dg}$, as part of the Cartan subgroup, is commonly admitted. To conclude, the compact manifold, and the supersymmetric need of an Euclidean $g_{\Ga\Gb}$,
do fit well together with the consideration of $U(1)^{\dg}$.}. Eventually, we will take $g_{\Ga\Gb}$ to be the identity, following the heterotic T-duality analogy and \cite{MS}.

This discussion raises another difficulty for a non-abelian gauge group: what should be chosen for $g_{\Ga\Gb}$? The previous relation (\ref{matchmet}) with
$\tr(t_{\Ga}t_{\Gb})$ would lead to a negative-definite metric, so it may not suit, in particular for a supersymmetric case. A possible answer comes from the heterotic T-duality. As discussed
in the Introduction, the metric $g_{\Ga\Gb}$ is related there to the symmetric part of the Cartan matrix of the gauge group\footnote{Note this matrix does
not exist in the abelian case.}. So this matrix could help, even if the way it would appear is not clear.

\section{Higher dimensional supersymmetry conditions}\label{secSUSY}

The search for supersymmetric (SUSY) flux solutions of ten-dimensional heterotic string has always been an important topic on the way to phenomenology.
 The conditions for
 finding such a vacuum are given by the annihilation of the supersymmetric variations of the fermions (the gravitino, dilatino and gaugino). When the
 ten-dimensional space-time is split into the warp product of a four-dimensional maximally symmetric space-time and a six-dimensional compact internal
 Riemannian manifold, the conditions for supersymmetry can be rephrased in terms of geometric quantities of the internal\footnote{The four-dimensional
 space-time is in addition constrained to be Minkowski and the warp factor vanishes (see also \cite{HLMM} for a derivation of this result).}
 manifold \cite{S, Hu}. We will give these SUSY conditions below. Following the spirit of the equivalence previously worked out, we propose in this section
 ``$\tD$-dimensional SUSY conditions'', which are equivalent to the internal SUSY conditions of heterotic string.

The bosonic theory introduced in $\tD$ dimensions is rather unlikely to be made supersymmetric. One could add fermions to it, but restricting
the highest spins to two, it is known that we would get $\tD \leq 11$. Since we would like $\dg\geq 1$, this theory could only be SUSY for
 $\dg=1$, which is rather restrictive. Therefore, our ``$\tD$-dimensional SUSY conditions'' will not be given by hypothetical higher dimensional
 supersymmetric variations of fermions, which are unlikely to exist. Instead, we will start from higher dimensional geometric conditions. Similarly to the
 e.o.m., these conditions will be the natural generalization of the internal (six-dimensional) heterotic conditions, in absence of gauge potentials. In addition,
we will show that they are equivalent to these six-dimensional SUSY conditions, once we use the ``heterotic ansatz''. Before giving
 these conditions, let us first introduce the necessary geometric ingredients.

\subsection{Setting the stage}

Our previous ``heterotic ansatz'' needs to be refined. The $D$-dimensional space (where we finally took $D=10$) needs to be split in a simple product of
 four-dimensional Minkowski times a manifold $\mmm$ of dimension denoted $d=D-4$. Eventually, $\mmm$ will correspond to the internal six-dimensional
 compact manifold. For now we only assume that $d$ is even, that this manifold is Riemannian, and that it admits an almost hermitian structure. In other words,
 this space has Euclidean signature and one can find there an almost complex structure we denote $I$. The corresponding $(1,0)$ and $(0,1)$ indices are
 denoted $\mu, \ov{\mu}$, and one can find an hermitian metric, denoted $g_{\mu\ov{\nu}}$. The fundamental form $J$ can then be defined on this space
 out of $I$ and $g$ (see appendix \ref{apcomplex} for our conventions). We denote a generic basis of one-forms on $\mmm$ by $(\d z^{\mu}, \d \ov{z^{\nu}})$.

Let us now consider the $\dg$-dimensional space. We assume as well that $\dg$ is even, and that the manifold $U(1)^{\dg}$ is Riemannian\footnote{This last restriction
is needed in order to have an hermitian metric. See section \ref{secnonab} for a discussion on this signature.}. Then, this torus on its own has a complex structure $I_g$ with $(1,0)$ and
 $(0,1)$ indices denoted $\al, \ov{\al}$, and an hermitian metric $g_{\al\ov{\be}}$. As discussed in appendix \ref{apcomplex}, the
 relation between its own real metric and the hermitian one is given by
\beq
\d s_{\dg}^2 = g_{\Ga\Gb} \d x^{\Ga} \d x^{\Gb}= 2  g_{\al\ov{\be}} \d z^{\al} \d \ov{z^{\be}} \ .
\eeq
Note then that one has $\ov{g_{\al\ov{\be}}}=g_{\ov{\al}\be}$. Let us now consider the connections $A^{\Ga}$. For $b,c$ real constants, if $\d z^{\al}=b \d x^{\Ga} + i c \d x^{\Gb}$, then
we define the associated complex connection as $A^{\al}=b A^{\Ga} + ic A^{\Gb}$, and $A^{\ov{\al}}=\ov{A^{\al}}$. More generally, we define them with the
same linear transformation which takes the real coordinates to the complex basis. This way, the one-forms $Z^{\al}=\d z^{\al} +A^{\al}$ are well-defined
 on the total space.

To preserve Lorentz invariance in four dimensions, we restrict all the fields to depend only on the $\mmm$ coordinates, and furthermore the connections only
 live on $\mmm$. Therefore, our ansatz for the $\tD$-dimensional space now looks like
\begin{center}
\begin{tabular}{cccc}
$U(1)^{\dg}$ & $\hookrightarrow$ & $\nnn$          &             \\
             &                   & $\downarrow$    &             \\
             &                   & $\mmm$ & $\times \quad$ $(D-d=4)$ Minkowski \\
             &                   & \multicolumn{2}{c}{$\underbrace{\phantom{....................................................}}\quad$}\\
             &                   & \multicolumn{2}{c}{$D$-dimensional space-time}
\end{tabular}
\end{center}
where $\nnn$ is the total bundle made of the fibration of the torus over $\mmm$. In addition, we ask for an almost complex structure defined
 on $\nnn$. To get one, we consider that $A^{\al}$, respectively $\ov{A^{\al}}$, are $(1,0)$-, respectively $(0,1)$-forms, on $\mmm$. This way we can talk of
 the $(1,0)$-form $Z^{\al}$ on $\nnn$. In other words, the compatibility of the two almost complex structures $I$ and $I_g$ is somehow required.\\

Given this refined ansatz for the $\tD$-dimensional space, we are now going to rewrite the $\tD$-dimensional fields using $(1,0)$- and $(0,1)$-forms.
Following (\ref{defcomplex}) and the previous definitions, the $\tD$-dimensional metric with the ``heterotic ansatz'' can be rewritten as
\beq
\d s_{\tD}^2=\d s_{{\rm Mink}}^2 + 2  g_{\mu\ov{\nu}} \d z^{\mu} \d \ov{z^{\nu}} + 2 g_{\al\ov{\be}}\ (\d z^{\al} +A^{\al}) (\d \ov{z^{\be}} +\ov{A^{\be}}) \ .
\eeq
Similarly, the fundamental two-form on $\nnn$ is given by $\JN=J+J_g$, where $J$ is the one on $\mmm$, and
\beq
J_g=i\ g_{\al\ov{\be}}\ (\d z^{\al} +A^{\al})\w (\d \ov{z^{\be}} +\ov{A^{\be}}) \ .
\eeq
We also introduce the maximal $(\frac{d+\dg}{2},0)$-form $\ON$ on $\nnn$
\beq
\ON= N \bigwedge_{\al=1}^{\frac{\dg}{2}} (\d z^{\al} +A^{\al}) \w \Omega \ ,
\eeq
where $\Omega$ is the one defined on $\mmm$, and $N$ is a normalization constant to be fixed with the volume (\ref{vol}). Similarly,
the $\tD$-dimensional $B$-field and $H$-flux with the ``heterotic ansatz'' can be rewritten as
\bea
\tB&=&B + B_g  + 2c \re(g_{\al\ov{\be}}\ A^{\al} \w \d \ov{z^{\be}}) \\
\tH&=&\d B + 2c \re(g_{\al\ov{\be}}\ \d A^{\al} \w \d \ov{z^{\be}}) + \frac{\ap}{4} {\rm CS}(\to) \ ,
\eea
where $B$ is now restricted to live only on $\mmm$. As before, we define
\beq
H= \d B - 2c \re(g_{\al\ov{\be}}\ \d A^{\al} \w \ov{A^{\be}}) + \frac{\ap}{4} {\rm CS}(\omega_+)  \ ,\  F^{\al}=\d A^{\al} \ ,\nn
\eeq
so that, using (\ref{apo+}), we get
\beq
\tH = H + 2c \re(g_{\al\ov{\be}}\ F^{\al} \w \ov{Z^{\be}}) + O(\appd) \ .
\eeq
Let us emphasize that the forms introduced here are just a rewriting, using $(1,0)$- and $(0,1)$-forms, of the forms previously defined
with the ``heterotic ansatz'' in section \ref{sechetans}. In particular, one can check that the $H$ just introduced corresponds to the heterotic $H$-flux\footnote{One has
$\re(g_{\al\ov{\be}}\ \d A^{\al} \w \ov{A^{\be}})=\frac{1}{2}\left(g_{\al\ov{\be}}\ \d A^{\al} \w \ov{A^{\be}} + g_{\ov{\al}\be}\ \d \ov{A^{\al}} \w A^{\be} \right)=
\frac{1}{2} g_{\Ga\Gb}\ \d A^{\Ga} \w A^{\Gb}$, which gives (\ref{defHanshet}).}, once one uses the matching formulas given in section \ref{secmatch}.

\subsection{Equivalence of the SUSY conditions}

For $d=6$, the SUSY conditions are given \cite{S, Hu} by
\bea
&& \d (e^{-2\phi} \Omega)=0 \label{susyhet1}\\
&& \d (e^{-2\phi} J^{\frac{d}{2}-1}) =0 \label{susyhet2}\\
&& i(\del-\ov{\del}) J= H \label{susyhet3}\\
&& \fff_{\mu\ov{\nu}}J^{\mu\ov{\nu}}=0 \Leftrightarrow \fff \w J^{\frac{d}{2}-1} =0 \label{susyhet4}\\
&& \fff_{\mu\nu}=\fff_{\ov{\mu}\ov{\nu}}=0 \ , \label{susyhet5}
\eea
where the power expansion is done with the wedge product. We are going to prove that the following (naturally generalized) conditions are equivalent to the
 previous set of conditions
\bea
&& \d (e^{-2\tp} \ON)=0 \label{gensusy1}\\
&& \d (e^{-2\tp} \JN^{\frac{d+\dg}{2}-1}) =0 \label{gensusy2}\\
&& i(\del-\ov{\del}) \JN= \tilde{H} \ . \label{gensusy3}
\eea
These are what we called the ``$\tD$-dimensional SUSY conditions''. The derivatives are here defined on the total space, but we write them as those on
 $\mmm$, since the fields are assumed to depend only on $\mmm$ coordinates.

We start with (\ref{gensusy1}). Since $A^{\al}$ is $(1,0)$ on $\mmm$, then $A^{\al} \w \Omega=0$, so we could write
$\ON=N \bigwedge_{\al=1}^{\frac{\dg}{2}} dz^{\al} \w \Omega$. Therefore,
\beq
\d (e^{-2\tp} \ON)=N e^{-2\varphi} \bigwedge_{\al=1}^{\frac{\dg}{2}} \d z^{\al} \w \d (e^{-2\phi} \Omega) \ ,
\eeq
and we get
\beq
\d (e^{-2\tp} \ON)= 0 \Leftrightarrow \d (e^{-2\phi} \Omega)=0 \ .
\eeq

Let us now consider (\ref{gensusy2}). We have
\beq
\JN^{\frac{d+\dg}{2}-1}=C_{\frac{d+\dg}{2}-1}^{\frac{d}{2}}\ J^{\frac{d}{2}} J_g^{\frac{\dg}{2}-1} +
 C_{\frac{d+\dg}{2}-1}^{\frac{\dg}{2}}\ J^{\frac{d}{2}-1} J_g^{\frac{\dg}{2}} \ , \label{dev}
\eeq
where the $C$s are the binomial coefficients. We have $\d(e^{-2\phi} J^{\frac{d}{2}})=0$ because $\phi$ depends only on $\mmm$ coordinates. Furthermore,
 $J^{\frac{d}{2}} \w \d(J_g^{\frac{\dg}{2}-1})=0$ because $\d(J_g^{\frac{\dg}{2}-1})$ produces a two-form on $\mmm$.
Therefore, only the second term in (\ref{dev}) contributes:
\bea
\d (e^{-2\tp} \JN^{\frac{d+\dg}{2}-1}) =0 &\Leftrightarrow& \d (e^{-2\phi} J^{\frac{d}{2}-1} J_g^{\frac{\dg}{2}} )=0 \\
&\Leftrightarrow& \d (e^{-2\phi} J^{\frac{d}{2}-1}) J_g^{\frac{\dg}{2}} + e^{-2\phi} J^{\frac{d}{2}-1} \d (J_g^{\frac{\dg}{2}} )=0 \ .\label{dev2}
\eea
In (\ref{dev2}), there is a single component proportional to $\bigwedge_{\al=1}^{\frac{\dg}{2}} \d z^{\al} \w \d \ov{z^{\al}} $ which comes
out of the first term. Out of it, we deduce
\beq \d (e^{-2\phi} J^{\frac{d}{2}-1})=0 \ .\eeq
The second term of (\ref{dev2}) then has to vanish. It is proportional to
\beq
J_g^{\frac{\dg}{2}-1}\w J^{\frac{d}{2}-1} \w g_{\al\ov{\be}}\ \left( \d \ov{z^{\be}} \w \d A^{\al} -\d z^{\al}\w \d \ov{A^{\be}}
 +\d(A^{\al} \w \ov{A^{\be}})\right) \ .
\eeq
The maximal forms living purely on $U(1)^{\dg}$ are a $(\frac{\dg}{2}, \frac{\dg}{2}-1)$- and $(\frac{\dg}{2}-1, \frac{\dg}{2})$-form. The annihilation of
these two terms leads respectively to
\beq
\forall \al,\be \ ,\  J^{\frac{d}{2}-1} \w \d \ov{A^{\be}} =0 \ , \ J^{\frac{d}{2}-1} \w \d A^{\al} =0 \ .
\eeq
It implies that the whole second term of (\ref{dev2}) vanishes. So finally, we get
\beq
\begin{tabular}{cc|l}
$\d (e^{-2\tp} \JN^{\frac{d+\dg}{2}-1}) =0$ & $\Leftrightarrow$ & $\d (e^{-2\phi} J^{\frac{d}{2}-1})=0$ ,\\
 & & $\forall \al\ ,\ F^{\al}_{\mu\ov{\nu}}J^{\mu\ov{\nu}}=F^{\ov{\al}}_{\mu\ov{\nu}}J^{\mu\ov{\nu}}=0$ .
\end{tabular}
\eeq

We finally consider (\ref{gensusy3}). One can compute using the definitions that
\bea
i (\del-\ov{\del}) \JN -\tilde{H}&=& i(\del-\ov{\del}) J -H + O(\appd) \label{susyHmatch}\\
&& - g_{\al\ov{\be}} \left[ \left((c+1) \ov{\del} \ov{A^{\be}}+(c-1) \del \ov{A^{\be}} \right) \w Z^{\al} + \left((c+1) \del A^{\al}+(c-1) \ov{\del} A^{\al}\right) \w \ov{Z^{\be}} \right] \ .\nn
\eea
Annihilating the whole expression and considering the only terms in $\d z^{\al}$ and $\d \ov{z^{\be}}$, one gets
\beq
(c+1) \ov{\del} \ov{A^{\be}}+(c-1) \del \ov{A^{\be}} = 0 \ ,\ (c+1) \del A^{\al}+(c-1) \ov{\del} A^{\al} = 0 \ .\nn
\eeq
The match with heterotic string worked out in section \ref{secmatch} lead us to choose $c=1$. We deduce here that $\forall \al$, $A^{\al}$ is antiholomorphic,
i.e. $\del A^{\al}=0$, or in other words
\beq
\forall \al,\  F^{\al}_{\mu\nu}=F^{\ov{\al}}_{\mu\nu}=F^{\al}_{\ov{\mu}\ov{\nu}}=F^{\ov{\al}}_{\ov{\mu}\ov{\nu}}=0 \ .
\eeq
The second line of (\ref{susyHmatch}) vanishes this way. We deduce that up to order $O(\appd)$ terms, one has
\beq
\begin{tabular}{cc|l}
$i (\del-\ov{\del}) \JN -\tilde{H} =0$ & $\Leftrightarrow$ & $i(\del-\ov{\del}) J -H=0$ ,\\
 & & $\forall \al\ ,\ F^{\al}_{\mu\nu}=F^{\ov{\al}}_{\mu\nu}=F^{\al}_{\ov{\mu}\ov{\nu}}=F^{\ov{\al}}_{\ov{\mu}\ov{\nu}}=0$ .
\end{tabular}
\eeq

Given the identifications to perform to recover heterotic string effective description at order $\ap$, as discussed in section \ref{secmatch}, we conclude
 that the ``$\tD$-dimensional SUSY conditions'' (\ref{gensusy1}), (\ref{gensusy2}) and (\ref{gensusy3}) are equivalent to the heterotic SUSY conditions
 (\ref{susyhet1}) to (\ref{susyhet4}) at order $\ap$, together with (\ref{susyhet5}) which is recovered up to $O(\ap)$ terms.\\

Let us make a final comment on these SUSY conditions. The heterotic SUSY conditions (\ref{susyhet1}) to (\ref{susyhet3}), expressed in terms of the $SU(3)$
structure forms $J$ and $\Omega$, were rewritten in \cite{AMP} using particular polyforms, which correspond in Generalized Complex Geometry (GCG) to pure
 spinors (this follows similar work done for type II supergravity in \cite{GMPT4, GMPT5}). One can
wonder if the same could be done for the conditions (\ref{gensusy1}) to (\ref{gensusy3}) with $\JN$ and $\ON$. If so, the corresponding polyforms could have an interesting
 interpretation as pure spinors on a bigger generalized tangent bundle, which would include the gauge fields. We will come back in section \ref{secTd} to
 such a GCG approach of heterotic string effective description.

\section{Solutions of heterotic string}\label{secsol}

In section \ref{seceq}, we showed that the equations of motion and the Bianchi identity of the (abelian bosonic) heterotic string
 at order $\ap$ were equivalent to those of the $\tD$-dimensional theory with the ``heterotic ansatz''. In this section, we discuss and illustrate an
 important consequence of this equivalence, related to the solutions of these equations.

Let us consider a solution of the $\tD$-dimensional theory, which does have the form of the ``heterotic ansatz''. It means
that some directions among the $\tD$ ones are circles fibered over a base. Initially, we proposed to match these directions with the gauge ones, and
the solution would then correspond to a heterotic string solution with non-trivial gauge fields. However, from the $\tD$-dimensional point of view, everything
is geometric, and there is nothing special about these circles, so we could as well consider them as a part of what becomes the space-time of heterotic string. In other words,
one could also understand this solution as a solution of heterotic string without gauge field, but with a space-time partly made of circles fibered over some
 base. The two heterotic string solutions are just related by an exchange of directions in the $\tD$-dimensional theory. Finally, one could also think of a mixed
solution: some of the fibered circles become geometric, and others give gauge fields. We will come back to this possibility. To conclude, one solution in
 the $\tD$-dimensional theory with the ``heterotic ansatz'' can give different solutions of heterotic string, with different geometries and gauge content.
 Those solutions are related by simple exchanges of directions in the $\tD$-dimensional theory.\\

We can illustrate this point explicitly with two solutions known in the literature. These solutions are supersymmetric and of the same type as those
 discussed in section \ref{secSUSY}: the ten-dimensional space-time is split in a four-dimensional Minkowski space-time times a six-dimensional compact
 manifold $\mmm$. The latter is given here by a
fibration of a two torus $T^2$ over a base $\bbb$, which is a conformal Calabi-Yau (CY) with a conformal factor related to the dilaton. The $T^2$ fibration is
 encoded in the two connections $A^{i=1,2}$ along the directions of coordinates $x^{i=1,2}$. We go to a basis where $\alpha=A^1+i A^2$ is a $(1,0)$-form on
 the base. The coordinate $z=x^1+i x^2$ is then taken as a holomorphic coordinate. For simplicity, we can further go to a basis where the real metric
of the $T^2$ is the identity. The metric, the fundamental two-form and the $(3,0)$-form of $\mmm$ are then given by
\beq
\d s^2= e^{2\phi} \d s_{\bbb}^2 + |\d z +\alpha|^2 \ , \ J=e^{2\phi} J_{\bbb} + \frac{i}{2} (\d z +\alpha)\w(\d \ov{z} +\ov{\alpha})
\ , \ \Omega= e^{2\phi} \omega_{\bbb} \w (\d z +\alpha) \ , \label{ansatzGP}
\eeq
where $J_{\bbb}$ and $\omega_{\bbb}$ are respectively the fundamental two-form and $(2,0)$-form of the Calabi-Yau $\bbb$. In addition, the dilaton is
 restricted to depend only on the base coordinates, and $F=\d \alpha$ has to satisfy
\beq
F\w J_{\bbb} = 0 \ , \ F\w \omega_{\bbb} = 0 \ .
\eeq
Then, one can verify that the SUSY conditions (\ref{susyhet1}) and (\ref{susyhet2}) are satisfied \cite{GP}.

Let us now take $\bbb=K3$, and consider two known solutions of heterotic string at order $\ap$, which are of the previous form.
 They both preserve $\nnn=2$ SUSY. For the first solution, the fibration is trivial,
meaning that $\alpha=0$, so $\mmm$ is the simple product of the $T^2$ and the conformal $K3$. In addition, it has non-trivial gauge fields: those are abelian
and each $\fff^{\alpha}$ is a $(1,1)$-form primitive on the base, so it satisfies the SUSY conditions (\ref{susyhet4}) and (\ref{susyhet5}). Finally,
there is no $B$-field so $H$ is only non-zero at order $\ap$. The second solution \cite{DRS, BD, KY} has $\alpha \neq 0$, so the fibration is non-trivial. In
 addition, there is no gauge field, but the $B$-field is non-zero\footnote{One can get this expression by following the dualities used to derive this solution
in \cite{DRS, BD}.}: $B=\re(\alpha \w \d \ov{z})= A^1 \w \d x^1 + A^2 \w \d x^2$. For this last solution to
 be $\nnn=2$, $F$ is restricted to be $(1,1)$ on the base. This implies in particular the holomorphicity of $\alpha$: $\del \alpha=0$. In both solutions,
 the dilaton is a priori non-trivial, and we take it to be the same for simplicity.

Let us give a few comments on these solutions. Solving the remaining SUSY condition (\ref{susyhet3}), or similarly the $H$ BI in absence of $NS5$ source is
 not trivial. In particular, the second solution (the non-K\"ahler one) was first discovered and studied in \cite{DRS, BD}
 by dualities, in some limit (the orbifold limit). But its rigorous existence was proven in theorems \cite{FY, BBFTY}, which in particular study non-trivially the
 existence of solutions to this BI. In addition, they impose topological constraints involving the gauge bundle \cite{LY, FY}.
As a consequence, there are some non-trivial topological restrictions on the choice of the base $\bbb$: the choice of $\bbb=K3$ and not
 $T^4$ in the previous solutions is crucial \cite{GP, FY, BBFTY}.

We also mention in a footnote\footnote{\label{footaphet}A general (physical) criticism
on these solutions is the mixture which occurs with $\ap$ order quantities; in particular some compactification cycles may end up being stringy because they
get a typical size of order $\apd$ (see for instance a remark in \cite{BS}). To illustrate this, one can compute for instance $\d H$. Using
 (\ref{ansatzGP}), and (\ref{susyhet3}), one gets
\beq
\d H =  -2i \partial \ov{\partial} (e^{2\phi}) \w J_{\bbb} + F \w \ov{F} \ .\label{dHsol}
\eeq
On the other hand, the BI (\ref{BI}) indicates that $\d H$ is of order $\ap$ (this remains true in the presence of a $NS5$ source). Therefore, the quantities
appearing in (\ref{dHsol}), in particular the curvature two-form $F$ or the connection one-forms $A^i$, could be of order $\apd$. Another possibility is
 that the quantities on the right-hand side of the BI (\ref{BI}), in particular the curvature two-forms, also have some $\ap$ dependence. In both cases,
 one could end up with cycles of size $\apd$, i.e. stringy.

Within the ten-dimensional theory defined at order $\ap$, such solutions are in principle allowed, and are consistent. Problems may
 occur when compactifying these solutions, or when considering them within the full string theory. However, we will not make such considerations here, and we only
use these solutions to illustrate our $\tD$-dimensional construction.} a discussion on the $\ap$ dependence in these solutions.\\

These two solutions have been shown in the literature to be related by various transformations. First, let us mention the ``K\"ahler/non-K\"ahler transition''
 \cite{BTY, Se}. Thanks to a chain of dualities and limits in moduli space, these two solutions
can be shown to arise from M-theory compactifications on $K3 \times K3$. The transition then consists in exchanging the two $K3$, in particular their
$(1,1)$-forms, which correspond in the heterotic setting either to $\fff$ or to $F$. Via a duality between this M-theory setting and type IIA on
 $X_3 \times S^1$, with $X_3$ a CY three-fold, this exchange of the two $K3$ (and so the transition) could also be seen as a mirror symmetry for $X_3$ \cite{A}.
 We can additionally mention that these two solutions were also related via a local $O(6,6+16)$ transformation (a twist) in \cite{AMP}. Finally, these solutions
were related by a heterotic T-duality \cite{EM, MSp}, which we will come back to.\\

Let us now rewrite these heterotic solutions within our $\tD$-dimensional theory with the ``heterotic ansatz'' (these vacua being by definition bosonic, their rewriting using the $\tD$-dimensional theory is possible). For simplicity, we use, on the
 $\dg$ part of the space, a basis where $g_{\Ga\Gb}$ becomes the identity (since the gauge group of these solutions is abelian, this can be done,
 as discussed in section \ref{secnonab}). Then, the first and second solution can be rewritten respectively as
\bea
&& \!\!\!\!\!\!\!\!\!\!\!\!\!\!\!\!\!\!\!\! \d s_{\tD}^2=\d s_{{\rm Mink}}^2 + e^{2\phi} \d s_{K3}^2 +\sum_{i=1,2} (\d x^i)^2
 + \sum_{\Ga=1 \dots \dg} (\d x^{\Ga}+ A^{\Ga})^2 \ ,\ \quad \tB= \sum_{\Ga=1 \dots \dg} A^{\Ga} \w \d x^{\Ga} + B_g \ , \label{sol1}\\
&& \nn\\
&& \!\!\!\!\!\!\!\!\!\!\!\!\!\!\!\!\!\!\!\! \d s_{\tD}^2=\d s_{{\rm Mink}}^2 + e^{2\phi} \d s_{K3}^2 +\sum_{i=1,2} (\d x^i +A^i)^2
 + \sum_{\Ga=1 \dots \dg} (\d x^{\Ga})^2 \ , \ \quad \tB= \sum_{i=1,2} A^{i} \w \d x^{i} + B_g \ . \label{sol2}
\eea
This $\tD$-dimensional rewriting allows us to illustrate the point discussed at the beginning of this section. Suppose we take in the first solution
 (\ref{sol1}) only two connections $A^{\Ga}$ to be a priori non-zero, and equate them\footnote{Note that having $A^{\Ga}=A^i$ is a priori possible since
$\fff^{\alpha}$ and $F$ have the same properties: they have been restricted to be $(1,1)$-forms, primitive on the base. However, doing so brings an
 explicit $\apd$ dependence in the $A^i$, given the formula for $A^{\Ga}$ (see sections \ref{secap} and \ref{secmatch}). Having the one-form $A^i$ being
 of order $\apd$ is actually plausible and consistent for these solutions: one can expect such a dependence by looking at the BI of $H$. This aspect may
 nevertheless lead to some critics when considering these solutions in a broader context. See a discussion on this point in footnote \ref{footaphet}.} to the $A^i$ of the second solution (\ref{sol2}). While the two
solutions are very different from the ten-dimensional point of view, they are then exactly the same from the $\tD$-dimensional point of view, up to an
 exchange of directions.

This exchange of directions should correspond to the ``K\"ahler/non-K\"ahler transition'' mentioned above. In particular, exchanging our circles should
 match with the exchange of the $K3$ in M-theory. Nevertheless, no relation between the $\fff^{\alpha}$ and $F$ of each solution is stated in \cite{BTY, Se},
 while here we take them equal. In section \ref{secothTd}, we will show that this exchange of directions can be encoded in a
 heterotic T-duality, different from the Buscher rules. Such a T-duality was used in \cite{EM}, while studying the global aspects of these solutions under
 this transformation. The technique used in \cite{MSp}, mentioned to be a heterotic T-duality, should also correspond to our transformation.\\

It is now tempting to consider solutions with both geometric connections (so a priori non-K\"ahler) and gauge fields\footnote{Such solutions may have been
obtained already in \cite{MSp}, where the solution generating technique developed leads to some non-K\"ahler solutions with non-zero gauge field.}. For instance, from the
second solution (\ref{sol2}), one could exchange only one of the $T^2$ directions with a circle of the gauge part. One would then get fields
of the form
\bea
&& \d s_{\tD}^2=\d s_{{\rm Mink}}^2 + e^{2\phi} \d s_{K3}^2 + (\d x^{i=1})^2 + (\d x^{i=2} +A^{i=2})^2
 + (\d x^{\Ga=1}+ A^{\Ga=1})^2 + \sum_{\Ga=2 \dots \dg} (\d x^{\Ga})^2 \ , \nn\\
&& \tB= A^{i=2} \w \d x^{i=2} +A^{\Ga=1} \w \d x^{\Ga=1}  + B_g \ . \label{sol3}
\eea
The result is only one circle non-trivially fibered in $\mmm$, and one in the gauge part. From the $\tD$-dimensional point, it is
 the same solution as before, so this set of fields must still satisfy the $\tD$-dimensional e.o.m., and, thanks to the equivalence, the e.o.m. of
 heterotic string as well. However, it is unlikely
that this solution would still satisfy the SUSY conditions, because it is not possible to have $(1,0)$ connection one-forms $\alpha$ and $A^{\alpha}$
anymore. This ``third solution'' would still satisfy the Bianchi identity, but as mentioned in the Introduction, we do not discuss here its integrated
 version. In particular, the global aspects could differ from one solution to the other, under this exchange of directions. The study of
 the global aspects in \cite{FY, EM} was only done for SUSY solutions. Therefore, we conclude that this third solution (\ref{sol3}) to the equations of
 motion remains for us only a conjectured solution to the full set of constraints.

\section{T-duality and Generalized Complex Geometry for heterotic string}\label{secTd}

As mentioned in the Introduction, considering the $\tD$-dimensional theory together with the ``heterotic ansatz'' was first inspired by the T-duality
 transformations of heterotic string. Therefore, performing heterotic T-duality in our $\tD$-dimensional theory is now very natural. This opens the
perspective to introduce a Generalized Complex Geometry (GCG) approach for heterotic string, which includes naturally the gauge fields. However, we will
mention a few difficulties for such a construction.

In this section, we first recall basics of T-duality and GCG. We then focus on heterotic T-duality and relate it to our $\tD$-dimensional construction.
As an application, we come back to the exchange of directions relating the K\"ahler and non-K\"ahler SUSY solutions, discussed in the previous section.
We show it can be encoded in a T-duality, which is not given by the Buscher rules. Finally, we discuss the results of Buscher T-dualities. One of them could
 lead to new non-geometric solutions, while another one leads us to a comparison with type IIB SUGRA solutions, and a discussion on a GCG approach in
 heterotic string.

\subsection{T-duality on the NSNS sector and Generalized Complex Geometry}\label{secGCG}

The T-duality group, acting on a NSNS configuration where fields are independent of $\di$ directions, is given by $O(\di,\di)$. It is often simpler, formulation
 wise, to embed this T-duality action in the bigger group $O(D,D)$ where $D$ is the dimension of the whole space. The action of this bigger group can
 always restricted to act non-trivially only in the $\di$ directions. This allows us to write things in terms of the whole fields, and not only for their
components on the $\di$ directions. We will use this formulation in the following, having in mind a possible restriction to the isometries directions.

One representation of interest of the $O(D,D)$ group is given by the set of the $2D \times 2D$ matrices $O$ which leave the matrix $\eta$ invariant,
 meaning, for the following matrices with $D \times D$ blocks,
\bea
&& \eta=\begin{pmatrix} 0_D & 1_D \\ 1_D & 0_D \end{pmatrix} \ , \ O=\begin{pmatrix} a & b \\ c & d \end{pmatrix} \nn\\
&& \begin{array}{cc|c}
 & & a^T c +c^T a =0_D\\
O^T \eta O =\eta & \Leftrightarrow & b^T d +d^T b = 0_D\\
 & & a^T d +c^T b = 1_D
\end{array}
\eea
Note that $O \in O(D,D) \Leftrightarrow O^T \in O(D,D)$. The action of this group on the NSNS fields can be encoded in different manners. For an element
 $\tilde{O} \in O(D,D)$, one can transform the metric and B-field by acting on the combination $E=g+B$ by the fractional linear transformation
\beq
E \rightarrow E'= (\tilde{a} E +\tilde{b})(\tilde{c} E +\tilde{d})^{-1} \ ,
 \ e^{\phi^\prime}=e^{\phi} \left( \frac{|g^\prime|}{|g|} \right)^{\frac{1}{4}} \ , \label{fraclin}
\eeq
and we also gave the transformation for the dilaton. One can recognize the new metric and $B$-field in $E'$ by looking at its symmetric and antisymmetric
 parts. There is however a more convenient way to work out this transformation. One can consider what is called the generalized metric ($2D \times 2D$ matrix)
\beq
{\cal H} = \begin{pmatrix}
         g - B g^{-1} B & B g^{-1} \\
         - g^{-1} B & g^{-1}
      \end{pmatrix}\ , \label{genmet}
\eeq
which transforms linearly under an $O(D,D)$ element $O$
\beq
\mathcal{H} \mapsto \mathcal{H'} = O^T \mathcal{H} O \ .\label{ODDH}
\eeq
This transformation reproduces the fractional linear transformation for $O=\tilde{O}^T$.

As an example of T-duality transformation, the $O(D,D)$ element reproducing the Buscher rules \cite{B} is given by
\beq
O_T=\left( \begin{array}{cc|cc}
0_n & & 1_n & \\
 & 1_{D-n} & & 0_{D-n} \\
\hline
1_n & & 0_n & \\
 & 0_{D-n} & & 1_{D-n}
\end{array} \right) \ , \label{TdBusch}
\eeq
where one performs the T-duality transformation along the top $n$ directions, $n \leq \di$.\\

Generalized Complex Geometry (GCG) is a mathematical framework developed by Hitchin and Gualtieri \cite{HG} in which the $O(D,D)$ action and the generalized
 metric ${\cal H}$ are natural considerations. For a review on the use of these mathematical tools in flux compactifications, see \cite{K}. Let us briefly introduce
 here a few concepts. For a $D$-dimensional manifold $M$, one considers the generalized tangent bundle $E$ given by the fibration of the
cotangent bundle over the tangent bundle
\beq
\begin{tabular}{ccc}
 $T^*M$&$\hookrightarrow$&$E$\\
  & &$\downarrow$\\
  & &$TM$
\end{tabular}
\eeq
Locally, it is just given by $TM \oplus T^*M$, so the sections, called generalized vectors, are given by the sum of a vector and a one-form
\beq
V=v+\xi = \begin{pmatrix} v \\ \xi
\end{pmatrix} \in TM \oplus T^*M  \, .
\eeq
The matrix $\eta$ then provides a natural metric to couple vectors and one-forms
\beq
V^T \eta V=\frac{1}{2} \begin{pmatrix} v & \xi
\end{pmatrix} \ \begin{pmatrix} 0_D & 1_D \\ 1_D & 0_D \end{pmatrix} \
 \begin{pmatrix} v \\ \xi
\end{pmatrix}= v^M \xi_M \ .
\eeq
This bilinear is left invariant by the $O(D,D)$ action, provided it acts linearly on the generalized vectors $V'=O^{-1} V$.

In this context, given a metric $g$ and a two-form $B$ living on $M$, one can show that the generalized metric ${\cal H}$ is a metric on the generalized
 tangent bundle $E$. Its $O(D,D)$ transformation is then given by (\ref{ODDH}). The T-duality action is then very natural in this context: roughly speaking,
it acts similarly to a rotation on this bigger space $E$. One can also introduce generalized vielbeins $\eee$.
 For a Riemannian manifold $M$, one can define the ordinary vielbeins $e$ as $e^T\ 1_D\ e =g$, so one would consider here $\eee$ such that
\beq
   {\cal H} = \eee ^T
      \begin{pmatrix} 1_D & 0_D \\ 0_D &  1_D
\end{pmatrix} \eee\ .
\eeq
There are several possible choices for these $\eee$, related by $O(2D)$ transformations on the left. For physical reasons, one should actually restrict
 this $O(2D)$ freedom to some $O(D) \times O(D)$ of a particular form (see for instance \cite{GMPW}). Here, we define the generalized vielbeins as
\beq
\eee =
\begin{pmatrix} e & 0_D \\
         - e^{-T} B &  e^{-T}
\end{pmatrix} \ . \label{genviel}
\eeq
The natural $O(D,D)$ action on these objects is then $\eee'=\eee O$. Nevertheless, a particular $O(D) \times O(D)$ on the left being possible, it turns out
 for the Buscher rules that one should rather act as \cite{GMPW}
\beq
\eee'=O_T \eee O_T \ . \label{OTaction}
\eeq

Let us finally mention that one can define (pure) spinors on $E$. We mentioned these spinors at the end of section \ref{secSUSY}. They have been used
 in type II SUGRA \cite{GMPT4, GMPT5} and in heterotic string \cite{AMP} to rephrase the SUSY conditions. They have also been used to reformulate
four-dimensional heterotic effective theory, following some previous work in type II SUGRA (see \cite{ALO} and references therein). These spinors can actually
 encode all the fields of the NSNS sector. By looking at the spinorial representation of the $O(D,D)$ group, one can perform the T-duality
 transformation equivalently on these spinors. The resulting spinors would then encode the T-dual fields.

\subsection{T-duality in heterotic string}

As discussed in the Introduction, one can extend for heterotic string the T-duality group to $O(\di,\di+\dg)$, where $\dg$ is the dimension of the Cartan
 subgroup of the gauge group. A way to work out the transformation of the gauge fields, in addition to that of the metric and $B$-field, is to consider the
pseudo metric (\ref{extg}) and $B$-field (\ref{extB}), and to act on the resulting $\tg+\tB$ with fractional linear transformation (\ref{fraclin})
\cite{GRV, SW, GR}. As mentioned in the previous section, it is however more convenient to act on a generalized metric and generalized vielbein. These
 objects, which already appeared before for T-duality on the NSNS sector alone, were somehow extended to $\di+\dg$ dimensions in \cite{MS}. More precisely,
 a generalized vielbein of size $(2\di +\dg)\times (2\di +\dg)$ was found out there. It has the same form
 as (\ref{genviel}) except that the last $\dg$ lines and columns are truncated. Its vielbein and $B$-field are in agreement\footnote{To be precise, as they
were working in the abelian case, there is of course a mismatch for $g_g$ and $B_g$. We will come back to the values they chose, and actually pick the same.} with the pseudo metric (\ref{extg}) and
 $B$-field (\ref{extB}). The reason to truncate the last lines and columns is that they want to act linearly with the $O(\di,\di+\dg)$ transformation.
But as mentioned in the Introduction, one can embed this $O(\di,\di+\dg)$ inside an $O(\di+\dg,\di+\dg)$ transformation, provided the transformation is
 forced to preserve the structure of the pseudo metric and $B$-field. One could therefore consider the full $2(\di +\dg)\times 2(\di +\dg)$ generalized
 vielbeins, or to simplify the formulation, $2(D +\dg)\times 2(D +\dg)$ generalized vielbeins, and then only act non-trivially on the desired components.
 Such generalized vielbeins were considered in \cite{AMP}, motivated by the study of local $O(D+\dg,D+\dg)$ transformations.

To summarize, one can perform heterotic T-duality by acting on $2(D +\dg)\times 2(D +\dg)$ generalized vielbeins and generalized metric, provided the metric
and $B$-field involved have the particular form of the pseudo fields (\ref{extg}) and (\ref{extB}), and that the $O(D+\dg,D+\dg)$ transformation preserves
 this form. In this paper, we considered a theory in $\tD=D+\dg$ dimensions where the metric and $B$-field had exactly the same form as the pseudo metric
and $B$-field. Therefore, when considering the associated $2(D +\dg)\times 2(D +\dg)$ generalized vielbeins and generalized metric, we consider exactly
the good objects on which to act with heterotic T-duality. This transformation is then very natural in our $\tD$-dimensional theory.

We give the $2(D +\dg)\times 2(D +\dg)$ generalized vielbein
\beq
\tilde{\eee}=\begin{pmatrix} \te & 0_{\tD} \\ - \te^{-T} \tB &  \te^{-T} \end{pmatrix} = \begin{pmatrix} e & 0 & 0 & 0 \\ e_g A & e_g & 0 & 0 \\
-e^{-T} (B + A^T g_g A) & e^{-T} A^T (B_g -g_g) & e^{-T} & -e^{-T}A^T \\ e_g A & -e_g^{-T} B_g & 0 & e_g^{-T} \end{pmatrix} \ ,\label{genvieltD}
\eeq
where we introduced $B_g$ and $e_g^T e_g =g_g$ of coefficients $B_{\Ga \Gb}$ and $g_{\Ga \Gb}$, and $A$ of coefficients $A^{\Ga}_M$
 (see (\ref{pseudomatrix})). We recall that the action to be performed is an $O(D+\dg,D+\dg)$ linear action, restricted in such
a way that it preserves the whole structure of $\tilde{\eee}$ (up to the $O(\tD) \times O(\tD)$ freedom), and leaves invariant $g_g$ (or even $e_g$) and $B_g$ (in other words,
 it can only transform $e$, $B$, and $A$). This way, the $O(D+\dg,D+\dg)$ is nothing but an embedding of the proper $O(D,D+\dg)$ heterotic T-duality group. In section \ref{secCcl}, we come back to the rewriting of the heterotic string effective action covariantly with respect to the heterotic T-duality, using the generalized metric associated to (\ref{genvieltD}).

In the following, we will choose $e_g=1_{\dg}$ and $B_g=0$. In the Introduction, the prescription was given to use the Cartan matrix of the gauge
 algebra. Here we consider the gauge group to be $U(1)^{\dg}$, for which no Cartan matrix is defined. We then follow the discussion of section
 \ref{secnonab}, where it was argued that $g_g$ could in that case be chosen to be the identity. These values for $e_g$ and $B_g$ were also chosen
 in \cite{MS} where the T-duality was also performed in the abelian case.

\subsection{Relating K\"ahler and non-K\"ahler solutions via T-duality}\label{secothTd}

In section \ref{secsol}, we argued that two SUSY solutions, one K\"ahler (\ref{sol1}) and the other non-K\"ahler (\ref{sol2}), are simply related by exchanging
some directions in the $\tD$-dimensional theory. We now show that this exchange of directions can be encoded in a heterotic T-duality\footnote{A
 T-duality relating such solutions has been considered already in \cite{EM, MSp}.}, which is not given by the Buscher rules. Note that in both solutions,
 nothing depends on the $U(1)^{\dg}$ nor on the $T^2$ coordinates, so T-dualities can be performed along both sets of directions.

Consider first a change of basis given by a $GL(\tD)$ matrix $P$ such that the one-forms (in basis $\bbb_1$) transform as $\d X'=P^{-1} \d X$. Then the
corresponding action on the vielbein\footnote{The action on the left of the vielbein can be understood as a change of basis for the local frame directions;
 in that case it is given by $P^{-1}$ and not $P^T$ because the coefficients of the vielbein matrix are given by one index up and one down, while those
 of the $B$-field and the metric are given by two indices down. The $P^{-1}$ on the left can also be interpreted as the $O(\tD)$ freedom (for
 Euclidean signature). As we will see, our $P \in O(\tD)$.} and $B$-field matrices is given
by $\te'= P^{-1} \te P$, $\tB'=P^T \tB P$. This change of basis can then be encoded in the $O(D+\dg,D+\dg)$ matrix $O_P$ acting on the generalized
 vielbein as
\beq
\tilde{\eee}'= O_P^{-1} \eee O_P =\begin{pmatrix} P^{-1} & 0_{\tD} \\ 0_{\tD} &  P^{T} \end{pmatrix}
\begin{pmatrix} \te & 0_{\tD} \\ - \te^{-T} \tB &  \te^{-T} \end{pmatrix} \begin{pmatrix} P & 0_{\tD} \\ 0_{\tD} &  P^{-T} \end{pmatrix} \ . \label{tdeee}
\eeq
While the action on the right can be interpreted as a proper T-duality, the action on the left can be understood as the $O(\tD) \times O(\tD)$ freedom previously
 discussed. Indeed, if $P \in O(\tD)$ as it will be the case here, then this action on the left has the allowed form \cite{GMPW}.

 Now we can choose this $P$ such that it reproduces the exchange of directions. We show this works explicitly for the vielbein. Given that $\tB'=P^T \tB P$,
 the exchange of directions will also be reproduced for the $B$-field\footnote{Note that having $B_g=0$ and $B_{T^2}=0$ also guarantees the exchange
 of the $B$-fields of the two solutions.}. Let us consider the following vielbein
\beq
\te=\begin{pmatrix} e_{\bbb} & 0 & 0 & 0 \\ e_{\fff} A_{\fff} & e_{\fff} & 0 & 0 \\ e_{gA} A_{g} & 0 & e_{gA} & 0 \\ 0 & 0 & 0 & e_{g0} \end{pmatrix} \ .
\eeq
We split the $\tD=D+\dg$ space-time as follows: there is a base $\bbb$ with vielbein $e_{\bbb}$ on which the connections $A_{\fff}$ and $A_g$ live, and over
 which are fibered the geometric fiber $\fff$ of vielbein $e_{\fff}$ and the gauge part. The latter, of dimension $\dg$ is split into a fibered part of
vielbein $e_{gA}$ and a free part $e_{g0}$. In the solutions considered in section \ref{secsol}, one has the base $\bbb$ made of four-dimensional Minkowski
 times the conformal $K3$. The geometric fiber is $\fff=T^2$, and we restricted ourselves to $e_{\fff}=1_2$. We argued in the previous section that $e_{gA}$ and
 $e_{g0}$ are also chosen to be the identity. Finally, the two solutions considered have only one non-trivial fibration: either $A_{\fff}=0$ in (\ref{sol1})
 or $A_{g}=0$ in (\ref{sol2}). In order to work out the exchange properly, we need the dimensions of the non-trivial fibrations to be the same:
 $d_{\fff}=d_{gA}$, and it equates $2$ in the solutions considered. Then we consider the following matrix
\beq
P=\begin{pmatrix} 1_{d_{\bbb}} & 0 & 0 & 0 \\ 0 & 0 & 1_{d_{\fff}} & 0 \\ 0 & 1_{d_{\fff}} & 0 & 0 \\ 0 & 0 & 0 & 1_{d_{g0}} \end{pmatrix} \ ,
 \ P=P^{-1}=P^T \in O(\tD) \ .
\eeq
One can easily check that it produces the exchange of directions as desired:
\beq
\te'=P^{-1} \te P = \begin{pmatrix} e_{\bbb} & 0 & 0 & 0 \\ e_{gA} A_{g} &  e_{gA} & 0 & 0 \\ e_{\fff} A_{\fff} & 0 & e_{\fff} & 0 \\ 0 & 0 & 0 & e_{g0} \end{pmatrix} \ .
\eeq
Then, given that $e_{\fff}=e_{gA}=1_{d_{\fff}}$, and if we take as argued in section \ref{secsol} the connections of the solutions to be the same, we clearly
 exchange the two solutions with this transformation. Since it does not change $e_g$ and $B_g$ in our solutions, this transformation encoded as in (\ref{tdeee})
is a good heterotic T-duality. Note that comparing $O_P$ with $O_T$ of (\ref{TdBusch}), one can easily see that this T-duality is not given by the Buscher
rules.\\

The conjectured solution (\ref{sol3}), which admits both a connection along $\fff$ and along the gauge circles, can clearly be obtained from one
of the other solutions by a similar transformation. One should just adapt slightly $P$ so that only part of the connections are exchanged.

\subsection{Buscher T-dualities, type IIB supergravity, and Generalized Complex Geometry}\label{secBusch}

Let us now discuss the T-duals of the solutions (\ref{sol1}) and (\ref{sol2}), using Buscher rules for the T-duality. This means one has to act
 as in (\ref{OTaction}) with the matrix $O_T$ given in (\ref{TdBusch}), on the generalized vielbeins (\ref{genvieltD}). We will compare the results with
those of type IIB solutions, and make related remarks on a GCG approach in heterotic string.

In order to simplify the discussion, let us first introduce the following well-known example, which has similarities with the heterotic solutions. We
 consider three circles along $x^{1,2,3}$ and a metric and $B$-field given by
\beq
\d s^2 = \frac{\im \rho(x^3)}{\im \tau(x^3)} |\d x^1 + \tau(x^3) \d x^2|^2 + (\d x^3)^2 \ , \ B=-\re\rho(x^3) \d x^1 \w \d x^2 \ , \label{toy0}
\eeq
which can be written as
\beq
g =  \frac{\im \rho}{\im \tau} \begin{pmatrix} 1 & \re \tau & 0 \\ \re \tau & |\tau|^2 & 0 \\ 0 & 0 & \frac{\im \tau}{\im \rho} \end{pmatrix} \ ,
 \ B= \begin{pmatrix} 0 & -\re \rho & 0 \\ \re \rho & 0 & 0 \\ 0 & 0 & 0 \end{pmatrix} \ .
\eeq
$\tau$ is the complex structure and $\im \rho$ is the volume. In that case, a (Buscher) T-duality along $x^1$ results in the exchange
 $\tau \leftrightarrow \rho$, while a T-duality along $x^2$ leads to the exchange $\tau \rightarrow -\frac{1}{\rho}$, $\rho \rightarrow -\frac{1}{\tau}$.

Let us now consider a particular case. We choose $\im \rho = \im \tau=1$, so that the circle along $x^1$ is simply fibered
 over the base, which is along $x^2$ and $x^3$. The fibration is given by a connection one-form $\re \tau (x^3) \d x^2$
\beq
\d s^2 = (\d x^1 + \re \tau(x^3) \d x^2)^2 + (\d x^2)^2  + (\d x^3)^2 \ , \ B=\re\rho(x^3) \d x^2 \w \d x^1 \ . \label{toy}
\eeq
This example has the same form as the heterotic solutions (\ref{sol1}) and (\ref{sol2}), provided that $\re \rho =\re \tau$. Indeed, the direction $x^1$
is the fiber direction, which corresponds either to the $U(1)^{\dg}$ or the $T^2$ directions. The direction $x^2$ corresponds to the directions of the $K3$
given by the one-forms $A^i$ or $A^{\Ga}$. We take $\re \rho =\re \tau$ because in the heterotic solutions, the $B$-field depends on the connections in this
way.\\

Let us first consider the T-duality along $x^2$. In the previous example (\ref{toy}), it is allowed since nothing depends on $x^2$ (at least with this gauge
 choice for the connection). We do not know whether the analogous situation can be found in the heterotic solutions (in particular, we do not know the
 $K3$ metric, which could depend on all $K3$ coordinates). Nevertheless, let us consider the case where such a T-duality is allowed. Then, this T-duality is
 known to lead to a non-geometric set-up with so-called $Q$-flux. The reason is the following: the diagonal metric element along the T-dualised direction
 will go through the ``radius inversion'', which results here in
\beq
g_{MM} \sim \frac{1}{|\rho|^2} \sim \frac{1}{1+(A^{\Ga}_M)^2} \ {\rm or} \ \frac{1}{1+(A^{i}_M)^2} \ ,
\eeq
according to the solution. Since the connection one-form is usually not globally defined, this metric element will not be single-valued. One could then
use a T-duality to patch it. This is the typical non-geometric situation with $Q$-flux: the geometry is only locally well-defined.

It would be interesting to investigate this possibility further. In particular, non-geometric solutions obtained as T-duals of a solution with non-trivial
gauge fields, like solution (\ref{sol1}), would be new with respect to the type II SUGRA examples. One could also investigate any possible relation with the
non-geometric\footnote{Let us mention that the pseudo fields (\ref{pseudomatrix}) have also been used to discuss non-geometry for heterotic string in
 \cite{FWW}.} solutions obtained in \cite{MMS}.\\

Let us now consider the T-duality along $x^1$, i.e. the fiber. In both heterotic solutions, nothing depends on the $U(1)^{\dg}$ nor on the $T^2$
 coordinates, so T-dualities can be performed along these fiber directions. By performing a T-duality along the $\dg$ directions for solution (\ref{sol1}),
 or along the $T^2$ for solution (\ref{sol2}), i.e. along the non-trivially fibered circles, one gets a surprising result: the T-dual generalized vielbeins
 remain totally invariant! We get
\beq
\tilde{\eee}'=O_T \tilde{\eee} O_T = \tilde{\eee} \ .
\eeq
In other words, the metrics, $B$-fields and gauge potentials are invariant\footnote{Of course, the fact we chose the identity metric for both the
 $U(1)^{\dg}$ and the $T^2$ part plays a role: one should normally get the inversion of radius, which is obviously not seen with the identity.}. By looking
at the simple example (\ref{toy}), one can actually understand this result. The T-duality along $x^1$ results in $\tau \leftrightarrow \rho$. This exchange
 clearly leaves the solution invariant, since we asked for $\im \rho = \im \tau$ and $\re \rho =\re \tau$. The presence, and structure of the $B$-field
in these solutions therefore plays a crucial role, especially in solution (\ref{sol1}) with non-trivial gauge field, where the ansatz for the $B$-field mixed
 component is surprisingly important.\\

Let us note that such a situation is unusual in type IIB SUGRA. There, the common $SU(3)$ structure SUSY solutions on a compact manifold
 either have a $B$-field but no connection (solutions of type B, as in \cite{GKP}, with a conformal Calabi-Yau, an imaginary self-dual three-flux,
 and D3-branes and O3-planes sourcing an $F_5$-flux), or have no $B$-field but a connection (solutions of type C, as for instance in \cite{KSTT, Sc}, on a
 twisted torus, with D5-branes and O5-planes sourcing an $F_3$-flux).  These two sets of solutions can be T-dual to each other. The T-duality is then said
 to exchange the $B$-field and the connection. This can be understood from the simple example (\ref{toy0}), with either $\re \rho$ or $\re \tau$ being zero,
 and the T-duality resulting in $\tau \leftrightarrow \rho$. Such a T-duality therefore (ex)changes the solutions in type IIB, while in heterotic string,
 the solutions remain invariant.

The classification of $SU(3)$ structure SUSY solutions of type IIB SUGRA \cite{G} also contains so-called type A solutions, which are similar to those of heterotic
 string, even if none is known on a compact manifold. Solutions of type A and type C are known to be S-duals. This S-duality provides another way to
 understand the difference between heterotic and type IIB SUSY solutions. Note this S-duality is also present in the chain of dualities which relates
 heterotic and type IIB string\footnote{Via an orbifold limit, one can identify type IIB with D9/D5 branes on $T^4/\mathbb{Z}_2$ with type I on $K3$, and the
 latter is S-dual to heterotic on $K3$ (see for instance \cite{DRS, BD}).}. Under S-duality, the $H$-flux is exchanged with the RR flux $F_3$. This can
 also be seen in the SUSY conditions for an $SU(3)$ structure. For a six-dimensional compact manifold, in the large volume limit and with $e^{\phi}=1$ for
 simplicity, type C solutions of type IIB SUGRA lead to the following SUSY conditions
\beq
\d (\Omega)=0 \ , \ \d (J\w J)=0 \ , \ \d (J)=-* F_3 \ ,
\eeq
while the SUSY conditions of heterotic string discussed in section \ref{secSUSY} (or type A solutions) can be rewritten as
\beq
\d (\Omega)=0 \ , \ \d (J\w J)=0 \ , \ \d (J)=-* H \ .
\eeq
These remarks lead us to the conclusion that in heterotic string, or at least for the solutions we considered, the $H$-flux plays the role of the $F_3$ in type IIB SUGRA. This is the reason for the qualitative differences of the solutions.\\

This different behavior of the $H$-flux between heterotic and type IIB solutions may have some importance for a GCG approach in heterotic string.
The $B$-field plays a particular role in GCG, as being responsible for the non-trivial fibration of the generalized tangent bundle.
Indeed, the $B$-field is similar to a connection there, as can be seen for instance in the generalized metric or vielbein. In type IIB SUGRA, the RR fields
on the contrary cannot be viewed directly in the (generalized) geometry. Therefore, if the $H$-flux in heterotic string acts like a RR flux in type IIB
SUGRA, one should then not include the $B$-field in a GCG construction in heterotic string. Instead, one should consider for instance a trivially fibered generalized tangent
 bundle. Working on the SUSY conditions (see end of sections \ref{secSUSY} and \ref{secGCG}), such a conclusion was already reached in \cite{AMP}. Indeed, it was noticed
 there that the pure spinors should not contain the $B$-field as they would have in type IIB SUGRA.

However, we discussed in this paper a generalized vielbein and metric that should be used for T-duality in heterotic string. Those objects did
 contain the $B$-field, and we mentioned that they were natural objects of GCG. Therefore, we conclude that a GCG construction in heterotic string is
 complicated by the fact that the pure spinor approach, and the generalized metric approach, seem not to be easily compatible.

\section{Final remarks}\label{secCcl}

The main motivations and results of this paper have been summarized in the Introduction. In this section, we would like to conclude with some remarks.

One of the main results of this paper is to prove the equivalence of the heterotic string effective action at order $\ap$ with a higher dimensional action,
provided one uses a particular ansatz for the higher dimensional space and fields. This higher dimensional action is very close to the NSNS standard action:
only an $\ap$ term differs. This leads us to two comments. First, it is known \cite{MS, DFT} that the NSNS action can be rewritten in an $O(D,D)$ covariant
way in terms of the dilaton and the generalized metric (\ref{genmet}). This implies here that up to this $\ap$ term, the heterotic string
effective action at order $\ap$ can be rewritten in an $O(D+\dg,D+\dg)$ covariant way in terms of the dilaton and the generalized metric related to
 (\ref{genvieltD}). This was one of the motivations for this paper, and we conclude here that this covariant rewriting can be performed as we have just described. The crucial step for it has been the incorporation of the gauge fields in the higher dimensional fields and geometry.

Secondly, if we choose $\dg=16$, as it should be for a flat ten-dimensional heterotic string, we have then shown that up to this $\ap$ term, the
bosonic string in $26$ dimensions and the (abelian bosonic) heterotic string in $10$ dimensions have equivalent effective descriptions, provided one can
 plug the ``heterotic ansatz'' in the bosonic string. This result may have interesting consequences. Note first that similar results seem to have been obtained from the world-sheet point of view. Indeed, heterotic string is said to be embedded in bosonic string either via a truncation \cite{Truncbos}, or as a Kaluza-Klein reduction \cite{KKbos}. It would be nice to study further the relation to these papers, as it looks like we obtain in our paper a (target space) effective theory derivation of such results, at order $\ap$. These world-sheet studies may also help for the non-abelian extension. Another consequence of such an embedding would be to use the bosonic string understanding of some phenomena, and translate it into the heterotic string. For instance, one could study mirror symmetry, or also non-geometry (see section \ref{secBusch} and \cite{FWW}) in heterotic string, using the bosonic string perspective.

In this paper, we have presented an equivalence of two theories in different dimensions, which should be understood as the rewriting of one into the other, as far as the action and various equations are concerned, provided the ``heterotic ansatz'' is used (see comments below (\ref{eq:introSUSY})). We have not discussed whether this equivalence corresponds to a proper compactification. In particular, the equality of the actions is shown as a computational result, and we have not performed any type of dimensional reduction. We leave this to future work, but let us present here a few elements related to this question. First, the ``heterotic ansatz'' specifies that no field depends on the $U(1)^{\dg}$ coordinates, as in a Kaluza-Klein reduction. Secondly, using this ansatz, the equations of motion are the same for both theories, so one could talk of a consistent truncation. These are interesting elements to argue for a dimensional reduction. However, one should study further the fact that the ``heterotic ansatz'' is not the most general ansatz for a reduction on a $\dg$-dimensional torus. Indeed, the off-diagonal components of the metric and $B$-field are not independent as they should a priori be. We know though that the heterotic string effective action is recovered at this cost. Actually, the fact these components are not independent is related to the chirality of heterotic string. One way to understand this is the argument given below (\ref{extB}) about heterotic T-duality: the fact the off-diagonal components of the metric and $B$-field are related in this particular way implies that the last $\dg$ lines of $\tg+\tB$ are totally fixed and should be left unchanged. This breaks the group $O(D+\dg,D+\dg)$ towards only the heterotic T-duality group $O(D,D+\dg)$. The asymmetry of the latter is exactly due to the chirality of heterotic string, so this particular ansatz for the off-diagonal components is related to this chirality. Coming back to the world-sheet descriptions \cite{GRV, SW, GR} should make this link more precise. To treat our rewriting as a proper dimensional reduction, one should then study whether this choice of the off-diagonal components corresponds to a truncation of some modes, which again could be given by the chirality constraint. We leave these questions to future work, and only consider the results of this paper as a rewriting.\\

An important aspect of the equivalence worked out is that the gauge fields are now on the same footing as the NSNS fields, since they are part of the
$\tD$-dimensional geometry. It would be nice to bring such a structure to type II SUGRA. There, the gauge fields enter the game in a different
 fashion: they appear through the branes actions. In most of the type II solutions with compact manifold, they are not taken into account. One reason
is that the negatively charged sources in type II SUGRA are often enough to satisfy the BI, and one does need to consider higher order $\ap$
 corrections, with which the gauge fields would arise. Finding a similar structure incorporating the gauge fields would though be interesting. In particular,
given that the heterotic solutions offer relations between gauge fields and $B$-fields, and following the discussion on S-duality made in the last section,
one may wonder whether in type II SUGRA a relation between the $C_2$ RR gauge potential and the gauge fields could be found.

Finally, the mathematical aspects related to the existence of solutions in heterotic string are often rather intricate, and we hope the equivalence
 worked out could help in this direction. Note as well the recent results at order $\ap$ on the conditions for the SUSY equations and the BI to be
 sufficient to guarantee a solution to the e.o.m.: see \cite{I, MSp}, and references therein. The equivalence might also help to understand better this
 result.

\section* {Acknowledgements}

I would like to thank M. Abou-Zeid, D. Cassani, J. Gray, S. Groot Nibbelink, E. Palti, A. Tomasiello, D. Tsimpis and P. Vaudrevange for useful discussions. I acknowledge support of the Alexander Humboldt Foundation.

\newpage

\appendix

\section{Conventions on forms and Riemannian geometry}\label{apconv}

In this appendix, we give our conventions on forms, and review some elements of Riemannian geometry, needed for our heterotic string computations. We also
derive some useful formulas.

\subsection{Forms and Hodge star}\label{apforms}

Our convention for a $p$-form $A$ on a generic basis of one-forms $\{\d x^{m}\}$ is
\beq
A=\frac{1}{p!}A_{m_1\dots m_p}\d x^{m_1}\w \dots \w \d x^{m_p} \ .
\eeq
Therefore, a wedged form given by the wedge product of $A$ and a $q$-form $B$ is defined as
\beq
\frac{1}{(p+q)!}(A\w B)_{m_1\dots m_{p+q}}=\frac{1}{p!q!}A_{[m_1\dots m_p} B_{m_{p+1}\dots m_{p+q}]} \ ,
\eeq
where the right-hand side indices are totally antisymmetrized. The contraction of a one-form $\xi$ with $A$ is given
by the following $(p-1)$-form
\beq
\iota_{\xi} A = \frac{1}{(p-1)!}\xi^{m_1} A_{m_1\dots m_p} \d x^{m_2} \w \dots \w \d x^{m_p} \ ,\label{contraction}
\eeq
where the index is raised by a given metric. Therefore, we get
\beq
\iota_{\xi} (A\w B) = (\iota_{\xi} A) \w B + (-1)^p\ A \w (\iota_{\xi} B)\ .
\eeq

The totally antisymmetric tensor $\epsilon$ is defined by $\epsilon_{m_0 \dots m_k}=+1/-1$ for
 $(m_0 \dots m_k)$ being any even/odd permutation of $(0 \dots k)$, and $0$ otherwise. Given a metric $g$ of determinant
 (of absolute value) $|g|$, on a $d$-dimensional space, our conventions for the Hodge star $*$ is then
\beq
*(\d x^{m_1} \w \dots \w \d x^{m_k})= \frac{\sqrt{|g|}}{(d-k) !} \ \epsilon^{m_1 \dots m_k\ m_{k+1} \dots m_d}\
g_{m_{k+1} n_{k+1}} \dots g_{m_d n_d}\ \d x^{n_{k+1}} \w \dots \w \d x^{n_d} \ , \label{Hodge*}
\eeq
One gets
\beq
**A= (-1)^{(d-p)p+s}\ A= (-1)^{(d-1)p+s}\ A \ ,
\eeq
where $s=0$, respectively $1$, for Euclidean, respectively Lorentzian, signature.

Let us consider that the basis of one-forms of the space can be split into two parts corresponding to two subspaces $\mmm$ and $\nnn$.
Consider that $A$ is given by forms living on $\mmm$ and $B$ by forms on $\nnn$. Consider finally that the metric is block diagonal in this basis of
 one-forms. Then one has
\beq
*(A\w B)=(-1)^{q(d_{\mmm}-p)}\ (*_{\mmm} A) \w (*_{\nnn} B) \ , \label{splithodge}
\eeq
where $*_{\mmm}$ is the Hodge star on the subspace $\mmm$ of dimension $d_{\mmm}$. This formula is valid provided the order of the forms
corresponds to the orientation, meaning that for $m_i$, respectively $n_i$, indices on $\mmm$, respectively $\nnn$, one has
$\epsilon_{m_1 \dots m_{d_{\mmm}} n_1 \dots n_{d_{\nnn}}}=\epsilon_{m_1 \dots m_{d_{\mmm}}} \epsilon_{n_1 \dots n_{d_{\nnn}}}$.

\subsection{Complex forms}\label{apcomplex}

We consider a $d$-dimensional Riemannian space with $d$ even and an almost hermitian structure. It means one can find an almost
complex structure $I$ defining locally $(1,0)$ and $(0,1)$ directions, denoted with indices $\mu$ and $\ov{\mu}$, and
in a local basis, $I_{\mu}^{\ \ \nu}=i \delta_{\mu}^{\ \ \nu}$, $I_{\ov{\mu}}^{\ \ \ov{\nu}}=-i \delta_{\ov{\mu}}^{\ \ \ov{\nu}}$.
In addition, there exists an hermitian metric in the local basis $g_{\mu\ov{\nu}}$. Out of these two objects, one can construct the
fundamental two-form $J$ by defining its coefficients as $J_{\mu\ov{\nu}}=i g_{\mu\ov{\nu}}$, $J_{\ov{\mu}\nu}=-i g_{\ov{\mu}\nu}$.

To be consistent with our tensor and form definitions on a generic one-form basis $\{\d x^m \}$, we have
\bea
&& J= \frac{1}{2} J_{mn} \d x^m \w \d x^n
= \frac{1}{2}(J_{\mu\ov{\nu}} \d z^{\mu} \w \d \ov{z^{\nu}} + J_{\ov{\mu} \nu} \d \ov{z^{\mu}} \w \d z^{\nu})
= J_{\mu\ov{\nu}} \d z^{\mu} \w \d \ov{z^{\nu}} = i g_{\mu\ov{\nu}} \d z^{\mu} \w \d \ov{z^{\nu}} \ ,\nn\\
&& \d s^2 = g_{mn} \d x^m \d x^n
= g_{\mu\ov{\nu}} \d z^{\mu} \d \ov{z^{\nu}} + g_{\ov{\mu}\nu} \d \ov{z^{\mu}} \d z^{\nu}
= 2  g_{\mu\ov{\nu}} \d z^{\mu} \d \ov{z^{\nu}} \ . \label{defcomplex}
\eea
As an example, let us define the following $(1,0)$-form in terms of two real forms $\d x^m$, and real constants $a,b$:
\beq
\d z^1= a\ \d x^1 + i b\ \d x^2 \ ,\ \d \ov{z^1}= a\ \d x^1 - i b\ \d x^2 \ ,
\eeq
where we can restrict ourselves to $a,b>0$. Then we have
\beq
\d s^2= g_{11} (\d x^1)^2 + g_{22} (\d x^2)^2 + \dots = 2 g_{1\ov{1}} |\d z^1|^2 + \dots
\eeq
Provided that $g_{12}=0$ (we can always choose the $\d x^m$ in such a way), and that the dots denote orthogonal directions,
we get $2 g_{1\ov{1}} = \frac{g_{11}}{a^2} = \frac{g_{22}}{b^2}$ \ .

The volume form is defined\footnote{It is implicit that the order $1 \dots d$ corresponds to the positive orientation. In formula (\ref{vol}), it is also
assumed that the definition of the $\d z^{\mu}$ and their order is such that this orientation is preserved.} as
 ${\rm vol}= \sqrt{|g|} \bigwedge_{m=1}^d \d x^m$. With our conventions, we then have
\beq
J^{\frac{d}{2}}=\bigwedge_{\mu=1}^{\frac{d}{2}} i g_{\mu\ov{\mu}} \d z^{\mu} \w \d \ov{z^{\mu}} = {\rm vol} \ ,\label{vol}
\eeq
where the term in the middle is valid for a diagonal metric (always reachable via a change of basis). To illustrate this formula,
we can use the previous example: $\d z^{1} \w \d \ov{z^{1}}=-2i ab\ \d x^1 \w \d x^2$, and therefore
$i g_{1\ov{1}} \d z^{1} \w \d \ov{z^{1}}= \sqrt{|g_{11}g_{22}|} \d x^1 \w \d x^2$. Finally, using this example, one can show that
\beq
*\d z^{\mu}= -i \d z^{\mu} \w {\rm vol}_{\bot \mu , \ov{\mu}} \ , \label{*dz}
\eeq
and its complex conjugation.

\subsection{Elements of Riemannian geometry}\label{apcovder}

In a coordinate basis, we write the vectors as $e_{m}=\del_m$ and the one-forms as $\d x^m$. In the
local frame (a priori a non-coordinate basis), we write them respectively as $e_a$ and $\theta^a$. More precisely, vielbeins $e^a_{\ m}$ are defined such
 that $g_{mn}=e^a_{\ m} e^b_{\ n} \eta_{ab}$, where $\eta$, the Minkowski metric, is used for Lorentzian signature, and
should be replaced by $\delta$ for Euclidean signature. From now on, we will use generically $\eta$ having in mind this
possible replacement. The inverse vielbein is denoted $e_a^{\ m}$: $e^a_{\ m} e_b^{\ n}= \delta^a_b \delta^n_m$. Then one
 defines $e_a=e_a^{\ m} e_m$ and $\theta^a=e^a_{\ m} \d x^m$.

For a tensor $t$, one has
\beq
t=t_{m_1 \dots m_l}^{n_1 \dots n_u} e_{n_1} \otimes \dots \otimes e_{n_u} \otimes \d x^{m_1} \otimes
 \dots \otimes \d x^{m_l}=t_{a_1 \dots a_l}^{b_1 \dots b_u} e_{b_1} \otimes \dots \otimes e_{b_u} \otimes
\theta^{a_1} \otimes \dots \otimes \theta^{a_l}\ .
\eeq
For a generic connection $\Gamma^m_{np}$, the covariant derivative of $t$ is defined as
\bea
\nabla_n (t)=\big(\del_n t_{m_1 \dots m_l}^{n_1 \dots n_u} \!\!\!\!\!\!\! && +\ \Gamma^{n_1}_{n k}
 t_{m_1 \dots m_l}^{k\ n_2 \dots n_u} +\ \Gamma^{n_2}_{n k} t_{m_1 \dots m_l}^{n_1\ k\ n_3 \dots n_u} + \dots  \nn\\
 && \!\!\!\!\!\!\!\!\!\!\!\!\!\!\!\!\!\!\!\!\!\!\!\!\!\!\!\!\!\!\!\!\!\!\! -\ \Gamma^k_{n m_1} t_{k\ m_2 \dots m_l}^{n_1 \dots n_u}
-\ \Gamma^k_{n m_2} t_{m_1\ k\ m_3 \dots m_l}^{n_1 \dots n_u} - \dots \big)
 \ e_{n_1} \otimes \dots \otimes e_{n_u} \otimes \d x^{m_1} \otimes \dots \otimes \d x^{m_l} \label{covder}
\eea
As a definition, one has $\nabla_n e_p= \Gamma^m_{np} e_m$ and $\nabla_n \d x^p= -\Gamma^p_{nm} \d x^m$.
Similarly one defines $\nabla_b e_c= \Gamma^a_{bc} e_a$, and then one gets
\beq
\Gamma^a_{bc} =e^a_{\ m} e_b^{\ n}  (\del_n e_c^{\ m} +\Gamma^m_{np} e_c^{\ p} ) \ . \label{gabc}
\eeq
As a consequence, the formula (\ref{covder}) is also valid with all indices being local frame indices ($a,b, \dots$), and where we define
$\del_a=e_a^{\ m}\del_m$. Given these definitions, one also has $\nabla_a=e_a^{\ m}\nabla_m$.

We introduce $f^a_{\ bc}$ (it corresponds to a structure constant when considering a Lie algebra)
\beq
f^a_{\ bc}=e^a_{\ m} \left(e_b^{\ n} \del_n e_c^{\ m}- e_c^{\ n} \del_n e_b^{\ m} \right) \ .
\eeq
It measures the difference between antisymmetrized connections in coordinate and non-coordinate basis. The torsion tensor
can be defined with it:
\beq
T^m_{\ np}=2\Gamma^m_{\ [np]}=e_a^{\ m}e^b_{\ n} e^c_{\ p}\left(2\Gamma^a_{\ [bc]} - f^a_{\ bc} \right)
=e_a^{\ m}e^b_{\ n}e^c_{\ p} T^a_{\ bc} \ .\label{torsion}
\eeq

The connection one-form is defined as $\omega^a_{\ b}=\Gamma^a_{cb}\ \theta^c$. The ``spin-connection'' is defined as its
coefficient in a coordinate basis: $\omega^a_{\ b m}=\Gamma^a_{cb} e^c_{\ m}$. The connection one-form satisfies the Cartan's
structure equations
\bea
&& \d \theta^a + \omega^a_{\ b} \w \theta^b = T^a = \frac{1}{2} T^a_{\ bc}\ \theta^b\w \theta^c \ ,\label{torsioncartan}\\
&& \d \omega^a_{\ b} + \omega^a_{\ c} \w  \omega^c_{\ b} =R^a_{\ b} = \frac{1}{2} R^a_{\ bcd}\ \theta^c\w \theta^d \ ,\label{Riem}
\eea
where the two lines give respectively the torsion two-form and the curvature two-form, out of the torsion and Riemann
tensors. We recall the torsion and Riemann tensors are always antisymmetric within their last two indices, whatever
connection is used. Note that (\ref{torsioncartan}) is obvious from (\ref{torsion}).

The exterior derivative is defined as $\d=(\d x^m \w) \del_m =(\theta^a \w) \del_a$. However, when acting on a form $t$, it can
be written with a covariant derivative involving a torsionless connection, thanks to the antisymmetrization:
\beq
(\d t)_{m_0 \dots m_l} = \del_{[m_0} t_{m_1 \dots m_l]}=\nabla_{[m_0} t_{m_1 \dots m_l]}
=e^{a_0}_{\ m_0} \dots e^{a_l}_{\ m_l} \nabla_{[a_0} t_{a_1 \dots a_l]} \ .\label{extder}
\eeq
Note that the last expression can be written with $\del_a$ and $f^a_{\ bc}$ since we consider in this formula $\nabla$ to be torsionless.
In addition, for the covariant derivative $\nabla_{\omega}$ associated to a generic connection $\omega$ and a
 two-form $t$, one has
\beq
\nabla_{\omega\ [m} t_{np]}=\del_{[m}t_{np]} + T^q_{\omega\ [mn} t_{p] q} \ , \label{torsionder}
\eeq
the formula being also valid with local frame indices. For a generic connection and a one-form $v=v_p \d x^p$, one can also show
\beq
2\nabla_{[m} \nabla_{n]} v_p = - T^r_{\ mn} \nabla_r v_p - R^q_{\ pmn} v_q \ . \label{doublecovder}
\eeq

The compatibility of the connection with the metric, meaning the metric being covariantly constant, is equivalent to the
 antisymmetry of the two free indices of the connection one-form: $\omega_{ab}=-\omega_{ba}$, where indices are raised
 and lowered with the appropriate metric. Then, out of (\ref{Riem}), one gets as well $R_{ab}=-R_{ba}$. We recall that the
Levi-Civita connection is the unique one compatible with the metric and torsionless. This is the connection used to derive Einstein equations.\\

Let us mention some possible confusions in notations. Here, we mean by $\nabla_n (t)_{m_1 \dots m_l}^{n_1 \dots n_u}$ the
coefficient of $\nabla_n (t)$ with indices $_{m_1 \dots m_l}^{n_1 \dots n_u}$ (formula (\ref{covder}) without the basis elements).
One has to be careful when the tensor has internal indices, like the curvature two-form. For instance, it is important to distinguish
 between $\nabla_n (R^a_{\ b})_{cd}$ and $\nabla_n (R)^a_{\ bcd}$: the first one is the derivative of
a two-tensor (here a two-form) while the second one is the derivative of a four-tensor (the Riemann tensor). A confusion can arise when one finds
a derivative without brackets, such as $\nabla_n R^a_{\ bcd}$. By convention, this would then mean the same as $\nabla_n (R)^a_{\ bcd}$, the derivative of
the Riemann tensor. An illustration of this distinction is given in the derivation of the following Bianchi identity.

By applying the exterior derivative on (\ref{Riem}), one can show
\beq
\d R^a_{\ b}= R^a_{\ c} \w \omega^c_{\ b} - \omega^a_{\ c} \w R^c_{\ b} \ , \label{BIRiem0}
\eeq
out of which one gets in components (lowering the $a$ to $e$, and where only $f,c,d$ are antisymmetrized)
\bea
&& \del_{[ f} (R_{eb})_{cd]} - \omega^k_{\ b[f} R_{ek |cd]} - \omega^k_{\ e[f} R_{kb |cd]}= 0\nn\\
&\Leftrightarrow& \nabla_{\omega\ [ f} (R_{eb})_{cd]} - \omega^k_{\ b[f} R_{ek |cd]} - \omega^k_{\ e[f} R_{kb |cd]}=  T^g_{\omega\ [fc} R_{eb |d] g}  \nn\\
&\Leftrightarrow& \nabla_{\omega\ [ f} R_{eb|cd]} = T^g_{\omega\ [fc} R_{eb |d] g} \nn\\
&\Leftrightarrow& \nabla_{\omega\ f} R_{ebcd} + \nabla_{\omega\ c} R_{ebdf} + \nabla_{\omega\ d} R_{ebfc} = 3\ T^g_{\omega\ [fc} R_{eb |d] g} \ . \label{BIRiem}
\eea
where we used (\ref{torsionder}). In the last line, we used the antisymmetry properties of the Riemann tensor. While in (\ref{BIRiem0}) the left-hand side
 is the covariant derivative of a two-form, the right-hand side provides the terms to make it a covariant derivative of
 a four-tensor, as we can see in the third line of (\ref{BIRiem}).

Going back to definitions, for a connection $\omega$ compatible with the metric, one can show that
\beq
\eta^{ef} \nabla_{\omega\ l} R_{\omega\ deaf} = \nabla_{\omega\ l} R_{\omega\ da} \ ,\label{derRic}
\eeq
where right-hand side is the covariant derivative of the Ricci tensor.

\subsection{Heterotic string connections}\label{aphetcon}

In heterotic string effective description, one introduces various connections. From now on, we denote $\omega^a_{\ b}$
the connection one-form corresponding to the Levi-Civita choice. One then introduces
\beq
\omega^a_{\pm \ b}=\omega^a_{\ b}\pm \frac{1}{2} H^a_{\ b} \ , \label{connecpm}
\eeq
where $H^a_{\ b}=e^a_{\ m} e_b^{\ n} g^{mp} H_{pnq} \d x^q$ is the one-form given by the $H$-flux coefficient. The latter is a tensor,
 so one goes from $a$ indices to $m$ indices with vielbein multiplications. Because of the
antisymmetry of $H$, one gets that $\omega_{\pm \ ab}$ is also antisymmetric in $a,b$, and so is the associated $R_{\pm\ ab}$.

Using (\ref{torsioncartan}), one can show that the torsion tensors associated to these connections are simply given by $H$:
\beq
T^a_{\epsilon=\pm1 \ bc}=-\epsilon\ H^a_{\ bc} \ . \label{torsionhet}
\eeq
We denote from now on by $\nabla$, respectively $\nabla_{\pm}$, the covariant derivative associated to $\omega^a_{\ b}$, respectively
 $\omega^a_{\pm \ b}$. As discussed in (\ref{extder}), when applied on a form $t$, the exterior derivative will often be considered
as $\d t=(\d x^m\w) \nabla_m t=(\theta^a\w) \nabla_a t$.\\

Several identities need to be derived. Note that in these derivations, we will not assume anything on the dimension of the space, nor on its signature.
 First, using (\ref{covder}) and (\ref{FormAB}), one can show for a two-form $t$
\beq
\forall m \ , \ \d x^m \w \d x^n \w \nabla_{\epsilon \ n}(*t) = \d x^m \w \d(*t) + \epsilon\ \d x^m \w t \w *H \ . \label{covderpmt}
\eeq
One can also give a formula for the Riemann and Ricci tensors. Using (\ref{Riem}), one gets
\bea
&& R_{\epsilon\ abcd}=R_{\ abcd} + \frac{1}{2} H_{ea[c} H^e_{\ d]b} + \epsilon \nabla_{[c} (H)_{d]ab} \ ,\label{Riemeps} \\
&& R_{\epsilon\ abcd} - R_{-\epsilon\ cdab} = 2 \epsilon \nabla_{[a} (H)_{bcd]} \ .\label{switchindices}
\eea
Using the antisymmetry properties, one can work out the associated Ricci tensor:
\beq
R^a_{\epsilon\ bae}=R_{\epsilon\ be}=R_{be} - \frac{1}{4} H_{bac}H_e^{\ ac} + \frac{\epsilon}{2} \left(\del_a H^a_{\ be}
+\omega^c_{\ ac}  H^a_{\ be} + \omega^c_{\ ba} H^a_{\ ec} - \omega^c_{\ ea} H^a_{\ bc} \right) \ , \nn
\eeq
which can be rewritten as
\beq
R_{\epsilon\ be}=R_{be} - \frac{1}{4} H_{bac}H_e^{\ ac} + \frac{\epsilon}{2} \eta_{bc} \eta_{ed} (-*)\left(
\theta^c\w \theta^d \w \d(*H) \right) \ , \label{Riceps}
\eeq
where one can work out the top form $\theta^c\w \theta^d \w \d(*H)$ using (\ref{covder}) and (\ref{extder}). The $(-*)$ is the notation for an
operator used to get the coefficient of this top form (see footnote \ref{foot-*} of appendix \ref{apom+} about it).

\section{Heterotic string equations of motion}\label{apheteom}

Given the conventions for heterotic string discussed in section \ref{sechetconv}, we are going to derive in details the equations
 of motion (e.o.m.) at order $\ap$ out of the action $S$ (\ref{action}), considered for a space of arbitrary dimension. For purposes of this paper, it is important
to note that the whole derivation does not depend on the dimension and the signature of the space
 (see in particular footnote \ref{foot-*} of appendix \ref{apom+}).

Deriving these e.o.m. at order $\ap$ is complicated by the dependence of the connection $\omega_+$ in various fields. In addition, $H$  does not depend only on the
 $B$-field but also on $\omega_+$ and on $\aaa$. Fortunately, the result is simplified thanks to an important lemma worked out in \cite{BR}. It states
 that the variation of the action with respect to $\omega_+$, which is of order $\ap$, is related to e.o.m. at order $\a$ (see (\ref{dSdo})).
 Therefore, this variation can be consistently discarded. We rederive here this lemma in details.

\subsection{First variations of the action}

The variation with respect to the dilaton gives us
\bea
&& 2\kappa^2 \frac{\delta S}{\delta \phi}= -2\sqrt{|g|} e^{-2\phi} \left(R -\frac{1}{2} |H|^2 + \frac{\ap}{4}
 (\tr (R_+^2)-\tr (\fff^2)) + 4(\nabla^2 \phi - |\d \phi|^2 ) + O(\app) \right) \ , \nn\\
&& E_{\phi\ 0}=R -\frac{1}{2} \left(|H|^2\right)_0 + 4(\nabla^2 \phi - |\d \phi|^2 )\ ,\
E_{\phi\ 1}=-\frac{1}{2} \left(|H|^2\right)_1 + \frac{\ap}{4} (\tr (R_+^2)-\tr (\fff^2)) \ ,
\eea
where $\left(\dots \right)_n$ denotes the part of $\dots$ being of order $\alpha^{\prime\ n}$. We introduce for later
 convenience the quantity to annihilate $E_{\phi\ n}$ at order $\alpha^{\prime\ n}$. For all the other fields, the derivation is more
subtle, because either $H$, or the connection $\omega_+$, depend on them. In view of the lemma previously discussed, we will
first write generically $\frac{\delta S}{\delta \omega_+}$, which will be studied later on. Therefore, the Einstein equation is given
 by\footnote{Let us recall briefly how to obtain the dilaton terms in the second line. Note that this derivation does not
 depend on the dimension. For a variation of the metric $g_{MN} \rightarrow g_{MN} + \delta g_{MN}$, one can show using the
 Levi-Civita connection that $\nabla_{M} \delta g_{N R} =\delta \Gamma^{S}_{MN} g_{S R} + \delta \Gamma^{S}_{MR} g_{S N}$.
 Then, one gets
\beq
g^{MN} \left( \nabla_{S} \delta \Gamma^{S}_{MN} - \nabla_{N} \delta \Gamma^{S}_{SM} \right)
= g^{MN} g^{RS} \nabla_{S} \left(\nabla_{M} \delta g_{N R} - \nabla_{R} \delta g_{M N} \right)\ .
\eeq
Using the Palatini identity and integration by parts, one gets the dilaton terms in the Einstein equation.}
\bea
2\kappa^2 \frac{\delta S}{\delta g^{MN}} &=& \sqrt{|g|} e^{-2\phi} \Bigg( \frac{\delta S}{\delta \omega_+}
 \frac{\delta \omega_+}{\delta g^{MN}} -\frac{g_{MN}}{2} \Big(R +4|\d\phi|^2- \frac{1}{2} |H|^2
 + \frac{\ap}{4} (\tr (R_+^2)-\tr (\fff^2)) \Big) \nn\\
&& \qquad \qquad \quad + R_{MN} - \frac{1}{2}\ \iota_{M} H \cdot \iota_{N} H +2 g_{MN}(2|\d \phi|^2-\nabla^2 \phi) + 2\nabla_{M}\nabla_{N} \phi \nn\\
&& \qquad \qquad \quad +\frac{\ap}{4} (\tr (\iota_{M} R_+ \cdot \iota_{N} R_+)-\tr (\iota_{M} \fff \cdot \iota_{N} \fff)) + O(\app) \Bigg) \ ,\nn
\eea
where we introduced for a $k$-form $A$ the notation $\iota_M A = \iota_{\d x^M} A$ out of (\ref{contraction}). Note in addition that
$\frac{\del |A|^2}{\del g^{MN}}= \iota_M A \cdot \iota_N A$. For later convenience, we introduce the following quantities:
\bea
E_{g\ 0, MN}&=&-\frac{g_{MN}}{2} E_{\phi\ 0} + R_{MN} - \frac{1}{2}\ \left(\iota_{M} H \cdot \iota_{N} H \right)_0 + 2\nabla_{M}\nabla_{N} \phi \ , \nn\\
E_{g\ 1, MN}&=&-\frac{g_{MN}}{2} E_{\phi\ 1} - \frac{1}{2}\ \left(\iota_{M} H \cdot \iota_{N} H \right)_1
+\frac{\ap}{4} (\tr (\iota_{M} R_+ \cdot \iota_{N} R_+)-\tr (\iota_{M} \fff \cdot \iota_{N} \fff)) \ , \nn\\
2\kappa^2 \frac{\delta S}{\delta g^{MN}} &=& \sqrt{|g|} e^{-2\phi} \Bigg( \frac{\delta S}{\delta \omega_+}
 \frac{\delta \omega_+}{\delta g^{MN}} + E_{g\ 0, MN} + E_{g\ 1, MN} + O(\app) \Bigg) \ .\nn
\eea
Similarly, we get for the variation with respect to $B$, up to a total derivative term,
\bea
&& 2\kappa^2 \delta S= \int \delta B \w \left[\d(e^{-2\phi} * H) + \frac{\delta S}{\delta \omega_+}
 \frac{\delta \omega_+}{\delta B} + O(\app) \right] \ , \nn\\
&& E_{B\ 0}= \d(e^{-2\phi} * (H)_0) \ , \ E_{B\ 1}= \d(e^{-2\phi} * (H)_1) \ .
\eea

Let us now vary the action with respect to $\omega_+$ and $\aaa$ only. The two variations are very similar, so we look at
 them together. Note that $\omega_+$ and $\aaa$ appear in the same manner through $H$, and through the $\ap$ term in the action.
Up to total derivative terms, we get
\bea
2\kappa^2 \delta S &=& -\frac{\ap}{4} \int \left(\delta \omega^a_{+\ b} \w \omega^b_{+\ a} + \tr(\delta A \w A) \right) \w \d(e^{-2\phi} * H) \nn\\
&& - \frac{\ap}{4} \int 2\ \tr(t_{\Ga} t_{\Gb})\ \delta \aaa^{\Ga} \w \left(\d(e^{-2\phi} * \fff^{\Gb})
+ e^{-2\phi} f^{\Gb}_{\ \ {\Gc}{\Gd}}\ \aaa^{\Gc} \w * \fff^{\Gd} - e^{-2\phi} \fff^{\Gb} \w *H \right) \nn\\
&& - \frac{\ap}{4} \int 2\ \delta \omega^a_{+\ b} \w \left(\d(e^{-2\phi} * R^b_{+\ a})
+ e^{-2\phi} \left(\omega^b_{+\ c}  \w * R^c_{+\ a} - \omega^c_{+\ a} \w * R^b_{+\ c} - R^b_{+\ a} \w *H \right) \right) \ .\nn
\eea
Using (\ref{covderpmt}), we rewrite the variation with respect to the gauge potential in the second line as
 $\delta \aaa^{\Ga} \w \d x^n \w \nabla_{-, \aaa\ n}(e^{-2\phi} * \fff^{\Gb})$, where the $\aaa$ subscript indicates a second covariantization
 with respect to the gauge field. The symmetry with the connection $\omega^a_{+\ b}$ allows us to rewrite the third line
in a similar manner. We will come back to its meaning in details. Therefore we get
\bea
2\kappa^2 \delta S &=& -\frac{\ap}{4} \int \left(\delta \omega^a_{+\ b} \w \omega^b_{+\ a} + \tr(\delta A \w A) \right) \w \d(e^{-2\phi} * H) \nn\\
&& - \frac{\ap}{4} \int 2\ \tr(t_{\Ga} t_{\Gb})\ \delta \aaa^{\Ga} \w \d x^n \w \nabla_{-, \aaa\ n}(e^{-2\phi} * \fff^{\Gb}) \nn\\
&& - \frac{\ap}{4} \int 2\ \delta \omega^a_{+\ b} \w \d x^n \w \nabla_{-, +\ n}(e^{-2\phi} * R^b_{+\ a}) \ . \label{varaction}
\eea

\subsection{The variation with respect to $\omega_+$}\label{apom+}

Let us rewrite the last line of (\ref{varaction}) as
\beq
 - \frac{\ap}{4} \int \d^{10} x \sqrt{|g|}\ 2\ \delta \omega^a_{+\ bc} (-*)
 \left(\theta^c \w \theta^f \w \nabla_{-, +\ f}(e^{-2\phi} * R^b_{+\ a}) \right) \ ,\label{lastline}
\eeq
where the $(-*)$ of a top form gives its coefficient\footnote{\label{foot-*}The minus is here symbolically taking into account the Lorentzian signature,
 but the whole derivation does actually not depend on the particular signature. The $(-*)$ should therefore be considered as a notation for the operator
 giving the coefficient, independently of the signature. In particular, this minus sign is never taken out of the parentheses.} (divided by $\sqrt{|g|}$
 for a coordinate basis). We are now interested in this coefficient that we denote $s$
\bea
s&=&(-*) \left(\theta^c \w \theta^f \w \nabla_{-, +\ f}(e^{-2\phi} * R^b_{+\ a})\right) = s_1 +s_2 \ , \nn\\
s_1&=& e^{-2\phi} (-*) \left(\theta^c \w \theta^f \w \nabla_{-, +\ f}(* R^b_{+\ a})\right) \ , \
s_2=(-*) \left(\theta^c \w \theta^f \nabla_{f}(e^{-2\phi}) \w * R^b_{+\ a}\right) \ .\nn
\eea
For a generic connection one-form $\Omega$ and a two-form $t$, one can prove using (\ref{covder})
\beq
\!\!\!\!\!\!\!\!\!\begin{array}{c|}
(-*)\left(\theta^c \w \theta^b \w \nabla_{\Omega\ b}(* t)\right) = \del_b t^{cb} + \Omega^c_{\ fb} t^{fb} - \Omega^b_{\ fb} t^{fc} \\
\nabla_{\Omega\ b}(t)_{de} = \del_b t_{de} - \Omega^f_{\ db} t_{fe} + \Omega^f_{\ eb} t_{fd}
\end{array}
\Rightarrow (-*)\left(\theta^c \w \theta^b \w \nabla_{\Omega\ b}(* t)\right) = \eta^{cd} \eta^{be} \nabla_{\Omega\ b} (t)_{de} \label{deriv2form}
\eeq
Furthermore, one can rewrite the $\omega_+$ covariantization in $s_1$ as
\beq
(-*)\left(\theta^c\w \left(\omega^b_{+\ f}  \w * R^f_{+\ a} - \omega^f_{+\ a} \w * R^b_{+\ f}\right) \right)
= \eta^{bl}\eta^{fe}\eta^{cd} \left(-\omega^k_{+\ lf} R_{+\ kade} -\omega^k_{+\ af} R_{+\ lkde} \right) \ .\nn
\eeq
Therefore, we get finally
\bea
s_1=e^{-2\phi} \eta^{bl}\eta^{fe}\eta^{cd} \Big(\del_f R_{+\ lade} \!\!\!\!\!\!\! &&  -\omega^k_{-\ df} R_{+\ lake} -\omega^k_{-\ ef} R_{+\ ladk} \nn\\
\!\!\!\!\!\!\! &&  -\omega^k_{+\ lf} R_{+\ kade} -\omega^k_{+\ af} R_{+\ lkde} \Big) \ ,
\eea
which is like the covariant derivative of the Riemann tensor, except that some covariantization are done with $\omega_+$ and
others with $\omega_-$. As for the Bianchi identity (\ref{BIRiem}), the terms in $\omega_+$ provide what is needed to go from
a covariant derivative on a two-form to one on a four-tensor. Now, we use (\ref{switchindices}), and in addition turn the two
 $\omega_+$ into $\omega_-$ in order to form a true covariant derivative on a four-tensor. We get
\bea
s_1=e^{-2\phi} \eta^{bl}\eta^{fe}\eta^{cd} \Big(\del_f R_{-\ dela} \!\!\!\!\!\!\! &&  -\omega^k_{-\ df} R_{-\ kela} -\omega^k_{-\ ef} R_{-\ dkla} \nn\\
\!\!\!\!\!\!\! &&  -\omega^k_{-\ lf} R_{-\ deka} -\omega^k_{-\ af} R_{-\ delk} \nn\\
\!\!\!\!\!\!\! && \!\!\!\!\!\!\!\!\!\!\!\!\!\!\!\!\!\!\!\!\!\!\!\!\!\!\!\! -H^k_{\ lf} R_{-\ deka} -H^k_{\ af} R_{-\ delk} \nn\\
\!\!\!\!\!\!\! && \!\!\!\!\!\!\!\!\!\!\!\!\!\!\!\!\!\!\!\!\!\!\!\!\!\!\!\!  +2 \nabla_{+,-\ f} \left(\nabla_{[l} (H)_{ade]} \right) \Big) \ ,
\eea
where $\nabla_{+,-\ f}$ means here that $\omega_+$ acts on the first two indices and $\omega_-$ on the last two.
Using (\ref{torsionhet}), we can rewrite this as
\bea
s_1=e^{-2\phi} \eta^{bl}\eta^{fe}\eta^{cd} \Big(\nabla_{-\ f} R_{-\ dela} \!\!\!\!\!\!\! && -3\ T^k_{-\ [fl} R_{-\ de|a]k} +H^k_{\ la} R_{-\ defk} \nn\\
\!\!\!\!\!\!\! && \!\!\!\!\!\!\!\!\!\!\!\!\!\!\!\!\!\!\!\!\!\!\!\!\!\!\!\!\!\!\!\!\!\!\! +2 \nabla_{+,-\ f} \left(\nabla_{[l} (H)_{ade]} \right) \Big) \ .
\eea
Using (\ref{BIRiem}) and (\ref{derRic}), we get the following, where the $R_-$ are now Ricci tensors
\beq
s_1=e^{-2\phi} \eta^{bl}\eta^{cd} \Big(2 \nabla_{-\ [a} R_{-\ d|l]}
 -H^k_{\ la} R_{-\ dk} +2\eta^{fe}\ \nabla_{+,-\ f} \left(\nabla_{[l} (H)_{ade]} \right) \Big)\ .
\eeq

We now turn to $s_2$. Using (\ref{switchindices}), we get
\beq
s_2=\eta^{bl}\eta^{fe}\eta^{cd} \nabla_{f} (e^{-2\phi}) R_{+\ lade}=-2 e^{-2\phi} \eta^{bl}\eta^{fe}\eta^{cd} \del_{f} (\phi)
 \left( R_{-\ dela}+ 2 \nabla_{[l} (H)_{ade]} \right) \ .
\eeq
Using (\ref{doublecovder}) and (\ref{torsionhet}), it becomes
\beq
s_2= -2 e^{-2\phi} \eta^{bl}\eta^{cd}  \left(2\nabla_{-\ [l} \nabla_{-\ a]} \del_{d} (\phi) + H^k_{\ la} \nabla_{-\ k}\del_{d} (\phi)
+ 2\eta^{fe}\del_{f} (\phi) \nabla_{[l} (H)_{ade]} \right) \ .
\eeq

Considering $s_2$ together with $s_1$, we finally get for $s$
\bea
s= e^{-2\phi} \eta^{bl}\eta^{cd} &\Bigg(& 2 \nabla_{-\ [a} \left( R_{-\ d|l]} +2 \nabla_{-\ l]} \del_{d} \phi \right)
 + H^k_{\ al} \left( R_{-\ dk} +2 \nabla_{-\ k}\del_{d} \phi \right) \nn\\
&& + 2\eta^{fe}e^{2\phi} \ \nabla_{+,-\ f} \left(e^{-2\phi}  (\d H)_{lade} \right) \Bigg) \ , \label{varfin}
\eea
where in the last line we used (\ref{extder}). Together with (\ref{lastline}) and (\ref{varaction}), we can reconstruct the variation of the action
 with respect to $\omega_+$. We obtain formula (\ref{dSdo}):
\bea
\!\!\!\!\!\!\!\! \frac{2\kappa^2}{\sqrt{|g|}} \frac{\delta S}{\delta \omega^a_{+\ bc}} &=& -\frac{\ap}{4} \Bigg[ (-*)\left(
\theta^c\w \omega^b_{+\ a} \w \d(e^{-2\phi} * H) \right) + 4 \eta^{bl}\eta^{cd}\eta^{fe} \ \nabla_{+,-\ f} \left(e^{-2\phi}  (\d H)_{lade} \right) \label{dSdoap}\\
&& + 2 e^{-2\phi} \eta^{bl}\eta^{cd} \Big( 2 \nabla_{-\ [a} \left( R_{-\ d|l]} +2 \nabla_{-\ l]} \del_{d} \phi \right)
 + H^k_{\ al} \left( R_{-\ dk} +2 \nabla_{-\ k}\del_{d} \phi \right) \Big) \Bigg] \ .\nn
\eea

Using (\ref{covder}) and (\ref{FormAB}), one can show that
\beq
\nabla_{\epsilon\ m}\del_n \phi= \nabla_{m}\del_n \phi + \frac{\epsilon}{2} g_{mq} g_{nr} (-*)
 \left(\d x^q\w \d x^r \w \d \phi \w *H \right) \ .
\eeq
Then, we reexpress the Ricci tensor (\ref{Riceps}) as
\beq
R_{\epsilon\ be}=R_{be} - \frac{1}{4} H_{bac}H_e^{\ ac} + \frac{\epsilon}{2} \eta_{bc} \eta_{ed} (-*)\left(
\theta^c\w \theta^d \w e^{2\phi} \d(e^{-2\phi} *H) \right) - 2 \left(\nabla_{\epsilon\ e}\del_b \phi- \nabla_{e}\del_b \phi \right) \ ,
\eeq
which can also be written with $m,n$ indices. Using the symmetry of the Ricci tensor $R_{be}$, we get the quantity
entering $s$, or the second line of (\ref{dSdoap}):
\beq
R_{-\ be} + 2 \nabla_{-\ e}\del_b \phi = E_{g\ 0, eb} + \frac{\eta_{eb}}{2} E_{\phi\ 0} - \frac{1}{2} \eta_{bc} \eta_{ed} (-*)\left(
\theta^c\w \theta^d \w e^{2\phi} E_{B\ 0} \right) + O(\ap) \ . \label{lasttrick}
\eeq
We are now able to conclude on the variation of the action with respect to $\omega_+$, and the e.o.m. at order $\ap$.

\subsection{Conclusion on the equations of motion}

According to (\ref{varaction}), the variation of the action with respect to $\omega_+$ and the gauge potential are of order
$\ap$. Therefore, if one solves the equations of motion order by order, one should first solve order $\a$ equations, meaning
$E_{\phi\ 0}$, $E_{g\ 0, MN}$ and $E_{B\ 0}$. Then one can consider the variations at order $\ap$. Let us first focus on the variation with
 respect to the gauge potential. Given the order $\a$ e.o.m. are satisfied, we read from (\ref{varaction}) that this variation leads to
\beq
e^{2\phi} \d(e^{-2\phi} * \fff) +  \aaa \w * \fff - * \fff \w \aaa - \fff \w (*H)_0 =0 + O(\ap)\ .
\eeq
Let us now consider the variation with respect to $\omega_+$, given in (\ref{dSdoap}). The term in $\d H$, which is of order $\ap$ (see
 below equation (\ref{BI})), then only contributes in this variation as $O(\app)$. In addition, thanks to (\ref{lasttrick}), we can see that the remaining terms in (\ref{dSdoap})
 vanish once the order $\a$ e.o.m. are satisfied. Therefore, the variation of the action at order $\ap$ with respect to $\omega_+$ can be consistently
discarded. This is the result of the lemma given in \cite{BR}, that we have rederived here.

To conclude, the equations of motion for the metric, $B$-field, and dilaton, are then corrected at order $\ap$ only with $E_{g\ 1, MN}$, $E_{B\ 1}$, and
$E_{\phi\ 1}$ respectively.

\section{Rewriting the $\tD$-dimensional theory}\label{apreduc}

In this appendix, we plug the ``heterotic ansatz'' detailed in section \ref{sechetans} into the $\tD$-dimensional theory, and rewrite its Einstein equation
 and $B$-field equation of motion (e.o.m.) accordingly. To do so, we compute first preliminary quantities. In the basis $\bbb_1$ written with the forms
 $(\d x^{M=0\dots D}, \d x^{\Ga=1 \dots \dg})$, one gets for the metric
$\d s_{\tD}^2= (g_{MN}+g_{\Ga\Gb}A_M^{\Ga} A_N^{\Gb}) \d x^M \d x^N + 2 g_{\Ga \Gb} A^{\Gb}_M \d x^M \d x^{\Ga} + g_{\Ga \Gb} \d x^{\Ga} \d x^{\Gb}$,
i.e. one can write the metric coefficients $\tg_{\tM\tN}$ (and its inverse $\tg^{\tM\tN}$) in this basis as
\bea
&& \tg_{MN}=g_{MN}+g_{\Ga\Gb}A_M^{\Ga} A_N^{\Gb} \ , \ \tg_{M\Ga}=g_{\Ga \Gb} A^{\Gb}_M \ , \ \tg_{\Ga \Gb}=g_{\Ga \Gb} \nn\\
&& \tg^{MN}=g^{MN} \ , \ \tg^{M\Ga}=-A^{\Ga}_N g^{NM} \ , \ \tg^{\Ga \Gb}=g^{\Ga \Gb}+ g^{MN} A_M^{\Ga} A_N^{\Gb} \ .\nn
\eea
This basis being a coordinate basis, we can use it to compute the connection
coefficients and the Ricci tensor. Nevertheless, we will come back later to the basis $\bbb_2$, more suited to rewrite the Einstein equation.
 Using the Levi-Civita connection both in $\tD$ and $D$ dimensions, one gets for its coefficients
\bea
&& \tG^{\tM}_{\Ga\Gb}=0 \ , \ \tG^{\Ga}_{\Gb M}= -\frac{1}{2} A^{\Ga}_N g^{NL} g_{\Gb\Gc} F^{\Gc}_{ML} \ ,
 \ \tG^M_{\Ga N}= \frac{1}{2} g^{ML} g_{\Ga\Gb} F^{\Gb}_{NL} \ , \nn\\
&& \tG^{\Ga}_{MN}=-A^{\Ga}_R (\Gamma^R_{MN} + g^{RL} g_{\Gb \Gc} A^{\Gb}_{(M} F^{\Gc}_{N)L}) + \del_{(M}A^{\Ga}_{N)} \ ,
 \ \tG^R_{MN}=\Gamma^R_{MN} + g^{RL} g_{\Gb \Gc} A^{\Gb}_{(M} F^{\Gc}_{N)L} \ , \nn
\eea
where we recall $F^{\Ga}_{MN}=2 \del_{[M}A^{\Ga}_{N]}$. Out of these\footnote{Note the useful formulas
\beq
\tG^M_{\Ga M}=0 \ , \ \tG^{\tM}_{N \tM}=\Gamma^M_{NM} \ .\label{relchris}
\eeq}, one can compute the components of the Ricci tensor $\tR_{\tM\tN}$:
\bea
\tR_{\Ga\Gb} &=& \frac{1}{4} g_{\Ga \Gc} g_{\Gb \Gd} F^{\Gc}_{MN} F^{\Gd\ MN} \nn\\
\tR_{\Ga M} &=& \frac{1}{2} g_{\Ga \Gb} \nabla_N (g^{NL} F^{\Gb}_{ML}) + \frac{1}{4} g_{\Ga \Gb} g_{\Gc \Gd} A^{\Gc}_M F^{\Gb}_{LN} F^{\Gd\ LN} \nn\\
\tR_{MN}&=& R_{MN}+ \frac{1}{4} g_{\Ga \Gb} g_{\Gc \Gd} F^{\Ga}_{PQ} F^{\Gc\ PQ} A^{\Gb}_M A^{\Gd}_N
 - g_{\Ga \Gb} \nabla_P(g^{PL}F^{\Ga}_{L(N}) A^{\Gb}_{M)} + \frac{1}{2} g_{\Ga \Gb} g^{LP} F^{\Ga}_{L(M} F^{\Gb}_{N)P}  \ .\nn
\eea
And finally, the scalar curvature is simply
\beq
\tR=R - \frac{1}{4} g_{\Ga \Gb} F^{\Ga}_{MN} F^{\Gb\ MN} \ .
\eeq

\subsection{The Einstein equation and the $H$-flux}

In the basis $\bbb_2$ given by $(\{\d x^M\}, \{\d x^\Ga+A^{\Ga}\})$, the metric is block diagonal:
 $\d s_{\tD}^2= g_{MN} \d x^M \d x^N + g_{\Ga \Gb} (\d x^{\Ga}+A^{\Ga}) (\d x^{\Gb}+A^{\Gb})$. As a consequence, the quantities entering the Einstein
 equations are simpler in this basis. Performing the change of basis on the Ricci tensor, we get the following components $\tR'_{\tM\tN}$ in $\bbb_2$:
\bea
\tR'_{MN}&=&\tR_{MN}-2\tR_{\Ga(M}A^{\Ga}_{N)}+\tR_{\Ga\Gb}A^{\Ga}_M A^{\Gb}_N \nn\\
&=& R_{MN} + \frac{1}{2} g_{\Ga \Gb} g^{LP} F^{\Ga}_{L(M} F^{\Gb}_{N)P}\\
\tR'_{\Ga M}&=&\tR_{\Ga M} - \tR_{\Ga \Gb} A^{\Gb}_M \nn\\
&=& \frac{1}{2} g_{\Ga \Gb} \nabla_N (g^{NL} F^{\Gb}_{ML}) \\
\tR'_{\Ga \Gb}&=&\tR_{\Ga \Gb} = \frac{1}{4} g_{\Ga \Gc} g_{\Gb \Gd} F^{\Gc}_{MN} F^{\Gd\ MN} \ .
\eea
Considering the same combinations of the dilaton term $\tna_{\tM} \del_{\tN} \tp$ (or equivalently bringing this two-tensor
to the basis $\bbb_2$ via the same transformation), it also gets simplified:
\bea
&& \tna_{M} \del_{N} \tp- 2\tna_{\Ga} \del_{(M} (\tp) A^{\Ga}_{N)} = \nabla_M \del_N \phi \\
&& \tna_{\Ga} \del_{M} \tp = -\frac{1}{2} g_{\Ga \Gb} g^{KL} F^{\Gb}_{MK} \del_L \phi \ .
\eea

The $H$-flux is in any case simpler when expressed in basis $\bbb_2$. One gets from (\ref{Hans})
\bea
&& \tH=H+c g_{\Ga \Gb} F^{\Ga} \w (\d x^{\Gb} + A^{\Gb}) \\
&& \Leftrightarrow \tH'_{MNP}= H_{MNP} \ , \ \tH'_{\Ga MN}= c g_{\Ga \Gb} F^{\Gb}_{MN} \ ,
\eea
up to order $\appd$ terms. Then, we deduce
\beq
|\tH|^2=|H|^2 + \frac{c^2}{2} g_{\Ga \Gb} F^{\Ga}_{MN} F^{\Gb\ MN} + O(\appd)\ .
\eeq
The Einstein equation in basis $\bbb_2$ involves the term in $\iota_{\tM} \tH' \cdot \iota_{\tN} \tH'$, for which one gets, up to order $\appd$ terms,
\bea
\iota_{M} \tH' \cdot \iota_{N} \tH' &=& \iota_{M} H \cdot \iota_{N} H+ c^2 g_{\Ga \Gb} \iota_{M} F^{\Ga} \cdot \iota_{N} F^{\Gb} \\
\iota_{M} \tH' \cdot \iota_{\Ga} \tH' &=& \frac{c}{2} g_{\Ga \Gb} F^{\Gb}_{NP} H_M^{\ \ NP} \\
\iota_{\Ga} \tH' \cdot \iota_{\Gb} \tH' &=& \frac{c^2}{2} g_{\Ga \Gc} g_{\Gb \Gd} F^{\Gc}_{MN} F^{\Gd\ MN} \ .
\eea

Finally, putting all the pieces together, the Einstein equation in $\tD$ dimensions (\ref{Einstein}) is equivalent to
\bea
&& \!\!\!\!\!\!\!\!\!\!\!\!\!\!\!\!\!\!\!\!\!\!\!\!\!\!\!\!\!\!\!\! R_{MN} - \frac{1}{2} \iota_{M} H \cdot \iota_{N} H  + 2 \nabla_M \del_N \phi
- \frac{c^2+1}{2} g_{\Ga \Gb} \iota_{M} F^{\Ga} \cdot \iota_{N} F^{\Gb} + \frac{\ap}{4} \tr (\iota_{M} R_+ \cdot \iota_{N} R_+) = 0 + O(\appd) \\
&& \!\!\!\!\!\!\!\!\!\!\!\!\!\!\!\!\!\!\!\!\!\!\!\!\!\!\!\!\!\!\!\! \frac{1}{2} g_{\Ga \Gb} \nabla_N (g^{NL} F^{\Gb}_{ML}) - \frac{c}{4} g_{\Ga \Gb} F^{\Gb}_{NP} H_M^{\ \ NP}
- g_{\Ga \Gb} g^{KL} F^{\Gb}_{MK} \del_L \phi = 0 + O(\appd)\label{FeomEins} \\
&& \!\!\!\!\!\!\!\!\!\!\!\!\!\!\!\!\!\!\!\!\!\!\!\!\!\!\!\!\!\!\!\! \frac{1-c^2}{4} g_{\Ga \Gc} g_{\Gb \Gd} F^{\Gc}_{MN} F^{\Gd\ MN}  = 0 + O(\appd)\ ,
\eea
where we used (\ref{Einstap}).

We can rewrite the off-diagonal equation (\ref{FeomEins}) in terms of forms. We first rewrite it equivalently as
\beq
g^{KM} g^{NL} \nabla_N (e^{-2\phi} F^{\Gb}_{ML}) - \frac{c}{2}e^{-2\phi}  F^{\Gb}_{NP} H^{KNP} = 0 + O(\appd) \ ,
\eeq
where we used the compatibility of the metric with respect to the Levi-Civita connection. Using (\ref{deriv2form}), (\ref{extder}) and (\ref{FormAB}), it
becomes
\bea
&& (-*_D) \d x^K \w \left( \d(e^{-2\phi} *_D F^{\Gb}) - c e^{-2\phi} F^{\Gb}\w *_D H \right) = 0 + O(\appd) \nn\\
&& \Leftrightarrow \d(e^{-2\phi} *_D F^{\Gb}) - c e^{-2\phi} F^{\Gb}\w *_D H = 0 + O(\appd) \ ,
\eea
where $(-*_D)$ should be understood as discussed in footnote \ref{foot-*} of appendix \ref{apom+}.

\subsection{The $B$-field equation of motion}

We are going to rewrite the $\tD$-dimensional $B$-field e.o.m. (\ref{Beom}) using the ``heterotic ansatz'', in particular the formula (\ref{Hans}) for $\tH$.
For simplicity, we introduce the notation $X^{\Ga}=\d x^{\Ga} + A^{\Ga}$. For the Hodge star, we are going to use the basis $\bbb_2$ since the metric is
 block diagonal there. To simplify the computation further, we go to the basis where $g_{\Ga\Gb}$ is diagonal. Being a constant symmetric matrix, the change
 of basis to perform only leads to a redefinition of the coordinates and connections via linear constant transformations. There is therefore no ambiguity
 in these redefinitions, and this change of basis is always possible. We do not change notations, so we consider $\tH$ to be given by
\beq
\tH=H+c g_{\Ga \Ga} F^{\Ga} \w X^{\Ga} \ ,
\eeq
up to order $\appd$ terms that we will not write for simplicity. Making use of (\ref{splithodge}), we get
\bea
*_{\tD} \tH &=& (*_{D} H) \w (*_{g} 1) + (-1)^D c g_{\Ga\Ga}\ (*_{D} F^{\Ga}) \w (*_{g} X^{\Ga})  \nn\\
&=& *_{D} H\ \frac{\sqrt{|g_g|}}{\dg !} \epsilon_{\Gb_1 \dots \Gb_{\dg}} \bigwedge_{\Gb=\Gb_1}^{\Gb_{\dg}} X^{\Gb}
+ (-1)^D c \ *_{D} F^{\Ga}\ \frac{\sqrt{|g_g|}}{(\dg-1) !} \epsilon_{\Ga\Gb_1 \dots \Gb_{\dg-1}} \bigwedge_{\Gb=\Gb_1}^{\Gb_{\dg-1}} X^{\Gb}  \ .
\eea
Then, after computation, we get
\bea
e^{2\varphi} \d(e^{-2\tp} *_{\tD} \tH)&=&\d(e^{-2\phi} *_{D} H)\ \frac{\sqrt{|g_g|}}{\dg !} \epsilon_{\Gb_1 \dots \Gb_{\dg}} \bigwedge_{\Gb=\Gb_1}^{\Gb_{\dg}} X^{\Gb}  \nn\\
&& +(-1)^D \left[-e^{-2\phi} *_{D} H\w F^{\Ga} + c \d\left(e^{-2\phi} *_{D} F^{\Ga} \right) \right]\
 \frac{\sqrt{|g_g|}}{(\dg-1) !} \epsilon_{\Ga\Gb_1 \dots \Gb_{\dg-1}} \bigwedge_{\Gb=\Gb_1}^{\Gb_{\dg-1}} X^{\Gb} \nn\\
&& +c e^{-2\phi} \frac{\sqrt{|g_g|}}{(\dg-2) !}\ *_{D} F^{\Ga}\w F^{\Gb}\ \epsilon_{\Ga\Gb\Gc_1 \dots \Gc_{\dg-2}} \bigwedge_{\Gc=\Gc_1}^{\Gc_{\dg-2}} X^{\Gc} \ .
\eea
First note that the last line automatically vanishes. Indeed, according to (\ref{FormAB}), $*_{D} F^{\Ga}\w F^{\Gb}$ is symmetric in $\Ga$ and $\Gb$ while
$\epsilon_{\Ga\Gb\Gc_1 \dots \Gc_{\dg-2}}$ is antisymmetric. Then, we can look at the first two lines using the basis $\bbb_1$, and consider the different
 degrees of the forms living on the $D$-dimensional space. Putting the whole expression to zero because of the equation of motion (\ref{Beom}), we get
conditions on these different forms. The smallest form living on the $D$-dimensional space is a $(D-2)$-form, obtained only out of the first line.
 From it we get:
\beq
\d(e^{-2\phi} *_{D} H)= 0 \ .
\eeq
The first line therefore vanishes. Then, the next smallest form is a $(D-1)$-form, coming out the second line. We get:
\beq
\forall \Ga\ , \ -e^{-2\phi} *_{D} H\w F^{\Ga} + c \d\left(e^{-2\phi} *_{D} F^{\Ga} \right) =0 \ .
\eeq
The second line then also vanishes, and nothing remains. To conclude, we get that the $B$-field equation of motion (\ref{Beom}) is equivalent to the two
 equations
\bea
&& \d(e^{-2\phi} *_{D} H)= 0 + O(\appd) \nn\\
&&  c \d\left(e^{-2\phi} *_{D} F^{\Ga} \right) - e^{-2\phi} F^{\Ga} \w  *_{D} H =0 + O(\appd) \ , \nn
\eea
where we put back the proper $\ap$ neglected terms. Note that the last equation is valid for any $F^{\Ga}$ defined in the basis where $g_{\Ga\Gb}$ is diagonal.
We can go back to the initial basis by performing on these $F^{\Ga}$ the inverse constant linear transformation. The result is of course the same: the last
equation is therefore valid in full generality.


\newpage

\begin{thebibliography}{}

\bibitem{V} C. Vafa, \textit{Evidence for F theory}, \textit{Nucl. Phys. B} \textbf{469} (1996) 403 [hep-th/9602022]

\bibitem{GPR} A. Giveon, M. Porrati, E. Rabinovici, \textit{Target Space Duality in String Theory}, \textit{Phys. Rept.} {\bf 244} (1994) 77 [hep-th/9401139]

\bibitem{T} D. C. Thompson, \textit{T-duality Invariant Approaches to String Theory}, [arXiv:1012.4393]

\bibitem{HG} N. Hitchin, \textit{Generalized Calabi-Yau manifolds}, \textit{Quart. J. Math. Oxford Ser.} \textbf{54} (2003) 281 [math.DG/0209099]\\
M. Gualtieri, \textit{Generalized complex geometry}, \textit{Oxford University DPhil thesis}, [math.DG/0401221]

\bibitem{GMPW} M. Gra\~na, R. Minasian, M. Petrini, D. Waldram, \textit{T-duality, Generalized Geometry and Non-Geometric
Backgrounds}, \textit{JHEP} \textbf{04} (2009) 075 [arXiv:0807.4527]

\bibitem{K} P. Koerber, \textit{Lectures on Generalized Complex Geometry for Physicists}, \textit{Fortsch. Phys.} \textbf{59} (2011) 169 [arXiv:1006.1536]

\bibitem{HuD} C. M. Hull, \textit{A Geometry for non-geometric string backgrounds}, \textit{JHEP} \textbf{10} (2005) 065 [hep-th/0406102]

\bibitem{DFT} O. Hohm, C. Hull, B. Zwiebach, \textit{Generalized metric formulation of double field theory}, \textit{JHEP} \textbf{08} (2010)
 008 [arXiv:1006.4823]\\
O. Hohm, C. Hull, B. Zwiebach, \textit{Background independent action for double field theory}, \textit{JHEP} \textbf{07} (2010) 016 [arXiv:1003.5027]

\bibitem{N} K. S. Narain, \textit{New Heterotic String Theories in Uncompactified Dimensions < 10}, \textit{Phys. Lett. B} \textbf{169} (1986) 41

\bibitem{NSW} K. S. Narain, M. H. Sarmadi, E. Witten, \textit{A Note on Toroidal Compactification of Heterotic String Theory}, \textit{Nucl. Phys. B}
\textbf{279} (1987) 369

\bibitem{GRV} A. Giveon, E. Rabinovici, G. Veneziano, \textit{Duality in String Background Space}, \textit{Nucl. Phys. B} \textbf{322} (1989) 167

\bibitem{SW} A. D. Shapere, F. Wilczek, \textit{Selfdual Models with Theta Terms}, \textit{Nucl. Phys. B} \textbf{320} (1989) 669

\bibitem{GR} A. Giveon, M. Ro\v{c}ek, \textit{Generalized duality in curved string backgrounds}, \textit{Nucl. Phys. B} \textbf{380} (1992) 128 [hep-th/9112070]

\bibitem{HS} S. F. Hassan, A. Sen, \textit{Twisting classical solutions in heterotic string theory}, \textit{Nucl. Phys. B} \textbf{375} (1992) 103
[hep-th/9109038]

\bibitem{MS} J. Maharana, J. H. Schwarz, \textit{Noncompact symmetries in string theory}, \textit{Nucl. Phys. B} \textbf{390} (1993) 3 [hep-th/9207016]

\bibitem{EGRS} S. Elitzur, E. Gross, E. Rabinovici, N. Seiberg, \textit{Aspects of bosonization in string theory}, \textit{Nucl. Phys. B} \textbf{283} (1987) 413

\bibitem{BR} E. A. Bergshoeff, M. de Roo, \textit{The Quartic Effective Action Of The Heterotic String And Supersymmetry}, \textit{Nucl. Phys. B}
 \textbf{328} (1989) 439

\bibitem{GP} E. Goldstein, S. Prokushkin, \textit{Geometric model for complex non-Kaehler manifolds with SU(3) structure}, \textit{Commun. Math. Phys.} {\bf 251} (2004) 65 [hep-th/0212307]

\bibitem{FY} J-X. Fu, S-T. Yau, \textit{The Theory of superstring with flux on non-Kahler manifolds and the complex Monge-Ampere equation}, \textit{J. Diff. Geom.}
\textbf{78} (2009) 369 [hep-th/0604063]

\bibitem{EM} J. Evslin, R. Minasian, \textit{Topology Change from (Heterotic) Narain T-Duality}, \textit{Nucl. Phys. B} \textbf{820} (2009) 213 [arXiv:0811.3866]

\bibitem{S} A. Strominger, \textit{Superstrings with torsion}, \textit{Nucl. Phys. B} \textbf{274} (1986) 253

\bibitem{Hu} C. M. Hull, \textit{Compactifications of the heterotic superstring}, \textit{Phys. Lett. B} \textbf{178} (1986) 357

\bibitem{DRS} K. Dasgupta, G. Rajesh, S. Sethi, \textit{M Theory, Orientifolds and $G$-flux}, \textit{JHEP} \textbf{08} (1999) 023 [hep-th/9908088]

\bibitem{BD} K. Becker, K. Dasgupta, \textit{Heterotic Strings with Torsion}, \textit{JHEP} \textbf{11} (2002) 006 [hep-th/0209077]

\bibitem{BBFTY} K. Becker, M. Becker, J-X. Fu, L-S. Tseng, S-T. Yau, \textit{Anomaly Cancellation and Smooth Non-K\"ahler Solutions in Heterotic String Theory}, \textit{Nucl. Phys. B} {\bf 751} (2006) 108 [hep-th/0604137]

\bibitem{BTY} M. Becker, L-S. Tseng, S-T. Yau, \textit{Heterotic K\"ahler/non-K\"ahler Transitions}, [arXiv:0706.4290]

\bibitem{Se} S. Sethi, \textit{A Note on Heterotic Dualities via M-theory}, \textit{Phys. Lett. B} {\bf 659} (2008) 385 [arXiv:0707.0295]

\bibitem{A} P. S. Aspinwall, \textit{An analysis of fluxes by duality}, [hep-th/0504036]

\bibitem{AMP} D. Andriot, R. Minasian, M. Petrini, \textit{Flux backgrounds from Twists}, \textit{JHEP} \textbf{12} (2009) 028 [arXiv:0903.0633]

\bibitem{MSp} D. Martelli, J. Sparks, \textit{Non-Kahler heterotic rotations}, [arXiv:1010.4031]

\bibitem{B} T. H. Buscher, \textit{A Symmetry of the String Background Field Equations}, \textit{Phys. Lett. B} \textbf{194} (1987) 59\\
T. H. Buscher, \textit{Path Integral Derivation of Quantum Duality in Nonlinear Sigma Models}, \textit{Phys. Lett. B} \textbf{201} (1988) 466

\bibitem{HLMM} J. Held, D. L\"ust, F. Marchesano, L. Martucci, \textit{DWSB in heterotic flux compactifications}, \textit{JHEP} \textbf{06} (2010) 090
[arXiv:1004.0867]

\bibitem{GMPT4} M. Gra\~na, R. Minasian, M. Petrini, A. Tomasiello, \textit{Supersymmetric Backgrounds from Generalized Calabi-Yau Manifolds}, \textit{JHEP} \textbf{08} (2004) 046 [hep-th/0406137]

\bibitem{GMPT5} M. Gra\~na, R. Minasian, M. Petrini, A. Tomasiello, \textit{Generalized structures of ${\cal N}=1$ vacua}, \textit{JHEP} \textbf{11} (2005) 020 [hep-th/0505212]

\bibitem{KY} S. Kim, P. Yi, \textit{A heterotic flux background and calibrated five-branes}, \textit{JHEP} \textbf{11} (2006) 040 [hep-th/0607091]

\bibitem{LY} J. Li, S-T. Yau, \textit{The existence of supersymmetric string theory with torsion}, [hep-th/0411136]

\bibitem{BS} K. Becker, S. Sethi, \textit{Torsional Heterotic Geometries}, \textit{Nucl. Phys. B} \textbf{820} (2009) 1 [arXiv:0903.3769]

\bibitem{ALO} L. Anguelova, F. Larsen, R. O'Connell, \textit{Heterotic Flux Attractors}, \textit{JHEP} \textbf{11} (2010) 010 [arXiv:1006.4981]

\bibitem{FWW} A. Flournoy, B. Wecht, B. Williams, \textit{Constructing nongeometric vacua in string theory}, \textit{Nucl. Phys. B} \textbf{706} (2005) 127
[hep-th/0404217]

\bibitem{MMS} J. McOrist, D. R. Morrison, S. Sethi, \textit{Geometries, Non-Geometries, and Fluxes}, [arXiv:1004.5447]

\bibitem{GKP} S. B. Giddings, S. Kachru, J. Polchinski, \textit{Hierarchies from Fluxes in String Compactifications}, \textit{Phys. Rev. D} \textbf{66}
 (2002) 106006 [hep-th/0105097]

\bibitem{KSTT} S. Kachru, M. B. Schulz, P. K. Tripathy, S. P. Trivedi, \textit{New supersymmetric string compactifications}, \textit{JHEP} \textbf{03} (2003) 061 [hep-th/0211182]

\bibitem{Sc} M. B. Schulz, \textit{Superstring orientifolds with torsion: O5 orientifolds of torus fibrations and their massless spectra}, \textit{Fortsch. Phys.} {\bf 52} (2004) 963 [hep-th/0406001]

\bibitem{G} M. Gra\~na, \textit{Flux compactifications in string theory: A comprehensive review}, \textit{Phys. Rept.} \textbf{423} (2006) 91 [hep-th/0509003]

\bibitem{Truncbos} A. Casher, F. Englert, H. Nicolai, A. Taormina, \textit{Consistent Superstrings as Solutions of the D=26 Bosonic String Theory}, \textit{Phys. Lett. B} \textbf{162} (1985) 121\\
F. Englert, H. Nicolai, A. Schellekens, \textit{Superstrings From Twentysix-dimensions}, \textit{Nucl. Phys. B} \textbf{274} (1986) 315

\bibitem{KKbos} M. J. Duff, B. E. W. Nilsson, C. N. Pope, \textit{Kaluza-Klein Approach To The Heterotic String}, \textit{Phys. Lett. B} \textbf{163} (1985) 343\\
M. J. Duff, B. E. W. Nilsson, N. P. Warner, C. N. Pope, \textit{Kaluza-Klein Approach To The Heterotic String. 2}, \textit{Phys. Lett. B} \textbf{171} (1986) 170\\
A. H. Chamseddine, M. J. Duff, B. E. W. Nilsson, C. N. Pope, D. A. Ross, \textit{SUPERSTRING sigma MODELS FROM BOSONIC ONES}, \textit{Phys. Lett. B} \textbf{193} (1987) 444

\bibitem{I} S. Ivanov, \textit{Heterotic supersymmetry, anomaly cancellation and equations of motion}, \textit{Phys. Lett. B} \textbf{685} (2010) 190
 [arXiv:0908.2927]


\end{thebibliography}
\end{document}